\documentclass[prd,onecolumn,preprint,nofootinbib]{revtex4}
\usepackage{graphicx}

\usepackage[plainpages=false, colorlinks=true, anchorcolor=blue, linkcolor=blue, citecolor=blue, bookmarks=false]{hyperref}
\usepackage{color}
\newcommand{\rthis}[1]{\textcolor{black}{#1}}
\usepackage{float}
\usepackage{amsfonts,amsmath,amssymb}
\usepackage{natbib}
\usepackage{array}
\usepackage{graphicx}
\usepackage{pdfpages}
\usepackage{booktabs}
\usepackage{hhline}
\usepackage[font=footnotesize]{caption}
\usepackage{multirow}
\usepackage{url}
\begin{document}
\newcolumntype{P}[1]{>{\centering\arraybackslash}p{#1}}
\pdfoutput=1
\newcommand{\jcap}{JCAP}
\newcommand{\araa}{Annual Review of Astron. and Astrophys.}
\newcommand{\apss}{Astrophysics and Space Sciences}
\newcommand{\aj}{Astron. J. }
\newcommand{\mnras}{MNRAS}
\newcommand{\physrep}{Physics Reports}
\newcommand{\apjl}{Astrophys. J. Lett.}
\newcommand{\apjs}{Astrophys. J. Suppl. Ser.}
\newcommand{\aap}{Astron. \& Astrophys.}
\newcommand{\pasa}{PASA}
\newcommand{\pasp}{PASP}
\renewcommand{\arraystretch}{2.5}
\title{An independent search for Jovian neutrinos using BOREXINO data}
\author{Yuva  Himanshu \surname{Pallam}$^1$} \altaffiliation{E-mail:f20220962@hyderabad.bits-pilani.ac.in}
\author{Shantanu \surname{Desai}$^2$ }
\altaffiliation{E-mail: shntn05@gmail.com}
\begin{abstract}
In a recent study, ~\citet{Saeed} found evidence for a 6\% flux contribution from Jupiter to the total rate time series data from the BOREXINO solar neutrino experiment, specifically during the time intervals 2019-2021 and 2011-2013.
 The significance of this detection was estimated to be around $2\sigma$. 
We reanalyze the BOREXINO data and independently confirm the Jovian signal with the same amplitude and significance as that obtained in ~\cite{Saeed}. However, using the same regression technique, we also find a spurious flux contribution from Venus and Saturn at $\sim 2\sigma$ significance, whereas prima facie one should not expect any signal from any other planet. We then implement  Bayesian model comparison  to ascertain whether the BOREXINO data contain an  additional contribution from Jupiter, Venus, or Saturn.  We find Bayes factors of less than five for  an additional contribution from Jupiter, and less than or close to  one for Venus and Saturn. This implies that the evidence for an additional contribution from Jupiter is very marginal. 
\end{abstract}

\affiliation{$^{1}$Department of Physics, BITS Pilani - Hyderabad Campus, Jawahar Nagar, Shameerpet Mandal, Hyderabad,  Telangana-500078, India}
\affiliation{$^{2}$Department  of Physics, IIT Hyderabad,  Kandi, Telangana-502284, India}
\maketitle
\section{Introduction}
In a recent work, ~\citet{Saeed} (A24 hereafter) found that the temporal variation in the $^7$Be solar neutrino signal from the BOREXINO experiment   exhibits a modulation, which they  attributed to a  contribution from Jupiter at about $2\sigma$ significance.  The flux from Jupiter has been estimated   to be about 6\% of the total observed solar neutrino signal. This Jovian contribution was also able to explain the suppression in the time variation of the signal  as well as  the lower value of the Earth's eccentricity. It was subsequently argued in A24 that this observed signal from Jupiter  is caused by capture and annihilation of dark matter WIMPs of masses less than 4 GeV in Jupiter's core. If this signal is independently confirmed, this would be the first evidence of neutrinos of MeV energy detected from any other solar system object other than the Sun. Although, a large number of detectors, starting with the Homestake experiment in 1960s~\cite{Homestake}  have measured the  solar neutrino flux, no other detector has found any signature  for Jovian neutrinos. In a similar vein, although neutrino detectors have been searching for signatures of dark matter capture and annihilation for  a long time~\cite{Desai04}, this would possibly be the first signature of WIMP annihilation to neutrinos  from any solar system object.

Therefore, given the profound implications of this result, it behooves us to  carry out an independent search for Jovian neutrinos in  the BOREXINO data. We therefore follow the same prescription as in A24 and supplement the analysis using Bayesian model comparison.  We then  also do a similar search from  some other  planets in order to compare the results with those  from  Jupiter, as a null test of the analysis method used. \rthis{One should not expect neutrinos from any other planet. If the same analysis technique shows  signals from other planets with similar significance as that observed from Jupiter, that would invalidate the results from Jupiter.}

This manuscript is structured as follows. We describe the BOREXINO data used for this analysis in Sect.~\ref{sec:data}. Our results using the total rate time series data and modulation data are presented in Sect.~\ref{sec:analysis} and Sect.~\ref{sec:modulation}, respectively. \rthis{The Jovian analysis using the full data can be found in Sect.~\ref{sec:fulldata}. The results from a frequentist analysis can be found in Sect.~\ref{sec:freq}.}
We conclude in Sect.~\ref{sec:conclusions}.
 
\section{Summary of A24}
\label{sec:data}
We now  briefly summarize the analyses carried out in A24 using the same notation. 
The BOREXINO Collaboration recently released 10 years of $^{7}$Be solar neutrino event rate time series data from December 2011 to October 2021 binned in one-month intervals~\cite{Borexino}.  The data set consists of total event rate time-series in  units of (cpd/100t), along with its uncertainty and also the estimated radioactive background.~\footnote{This data can be downloaded from \url{https://borex.lngs.infn.it/papers/articles/earths-orbital-parameters-from-solar-neutrinos/}} 

 In their first analysis, A24 modeled the BOREXINO rate time series $\mathcal{R}(t)$ using the following expression: 
 \begin{equation}
 \mathcal{R}(t) = \frac{\mathcal{R}_{sun}}{d^2_{sun}(t)} +  \frac{\mathcal{R}_{jup}}{d^2_{jup}(t)} + \mathcal{R}_B,
 \label{eq:R1}
 \end{equation}
 where $\mathcal{R}_{sun}$ and $\mathcal{R}_{jup}$ are proportional to the event rates induced by Sun and Jupiter, respectively, and $\mathcal{R}_B$ is the background contribution from radioactive sources. In Eq~\ref{eq:R1}, $d_{\mathrm{sun}}$ and $d_{\mathrm{jup}}$ denote the instantaneous distance to the Sun and Jupiter, respectively. 
 A24 assumed that $\mathcal{R}_{sun}$ and $\mathcal{R}_{jup}$ are constant. A24 then fit the data between October 2019 and October 2021 to Eq~\ref{eq:R1} using Bayesian regression, when the temporal variation of $\mathcal{R}_B$ was negligible~\cite{BorNature}.
  For Bayesian inference, uniform priors for $\mathcal{R}_{jup}$ and $\mathcal{R}_B$ were assumed, \rthis{since there is no prior theoretical model for a sigmal from Jupiter}. Both these priors are  given by $\in \mathcal{U} [0, 50]$. For $\mathcal{R}_{sun}$ two sets of priors were used: $\mathcal{U} [0, 50]$ and $\mathcal{N} (25, 2)$. A24 obtained the best-fit value of $\frac{\mathcal{R}_{jup}}{(5 \ {\rm AU})^2} = 1.5^{+0.7}_{-0.8} \ (\rm cpd/100t)$, which points to non-zero value at $2\sigma$.  

 Then a second analysis was done in A24, in which the event rate was written as a superposition of a trend and modulation term as follows:
 \begin{equation}
 \mathcal{R}(t) =  R_{tr} (t)+ \delta  \mathcal{R}(t),
 \label{eq:R2}
 \end{equation}
where $R_{tr} (t)$ corresponds to the trend of the data caused by the radioactive background, whose parametric form can be found in ~\cite{Borexino}, while its numerical value in each time bin is also provided in the BOREXINO public data release. After removing this trend of the data, A24 fitted the modulation  flux data $\delta  \mathcal{R}(t)$ to the following equation:
\begin{equation}
\delta \mathcal{R}(t) = \sum_i  \mathcal{R}_{i} \left[\frac{1}{d^2_{i}(t)} -   \frac{1}{T}\int_{exp}  \frac{dt}{d^2_{i}(t)}\right],
\label{eq:3}
\end{equation}
where the index $i$ represents the two objects considered, Sun and Jupiter and $T$ is the duration of the dataset analyzed.~\footnote{Equation (6) in A24 does not contain $T$, which is a typographical error (S. Ansarifard, private communication).} 
A24 found no contribution from Jupiter during 2015-2018. However, non-zero values were found during two distinct periods: December 2011-December 2013 and October 2019-October 2021 with best-fit values given by $\frac{\mathcal{R}_{jup}}{(5 \ {\rm AU})^2} = 1.6^{+0.8}_{-1.1} \ (\rm cpd/100t)$ and $\frac{\mathcal{R}_{jup}}{(5 \ {\rm AU})^2} = 1.7^{+0.8}_{-0.8} \ (\rm cpd/100t)$, respectively. Therefore, both methods yield a signal which is about 6\% of the total $^{7}$Be solar neutrino flux obsered.

Finally, A24 also repeated the same exercise on data from Sudbury Neutrino observatory  annual variation data from 1999-2003 in the energy range from 5-20 MeV~\cite{SNO}, but no signal was found.

\section{Analysis and Results}
\label{sec:analysis}
\begin{figure}[t]
    \centering
    \includegraphics[width=0.6\textwidth]{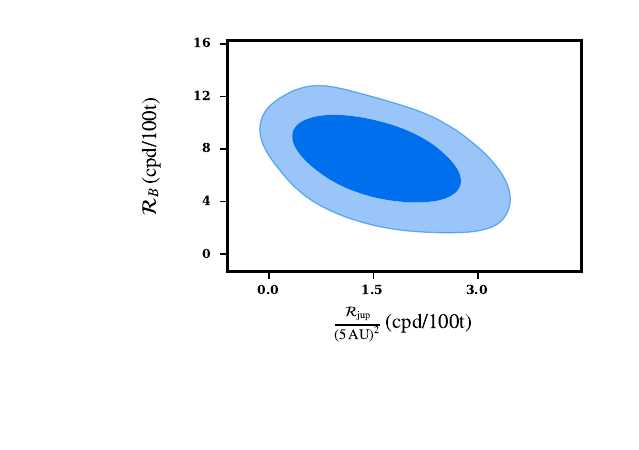}
    \caption{\label{fig:jup1} Marginalized 68\% and 95\% credible intervals for $\mathcal{R}_{\mathrm{B}}$ and  $\mathcal{R}_{\mathrm{jup}}$ after using uniform  prior on $\frac{\mathcal{R}_{\mathrm{jup}}}{(5\,\mathrm{AU})^2} \in \mathcal{U} [0, 4]$   and   normal prior on $\mathcal{R}_{\mathrm{sun}} \in \mathcal{N} (25,2)$ with units of (cpd/100t). For this plot, we have used DE442s ephemeris. The marginalized 68\% value for $\frac{\mathcal{R}_{\mathrm{jup}}}{(5\,\mathrm{AU})^2}$ is given by \( 1.53^{+0.78}_{-0.75} \) (cpd/100t).}
\end{figure}

\begin{figure}[htbp]
    \centering
    \includegraphics[width=0.9\textwidth]{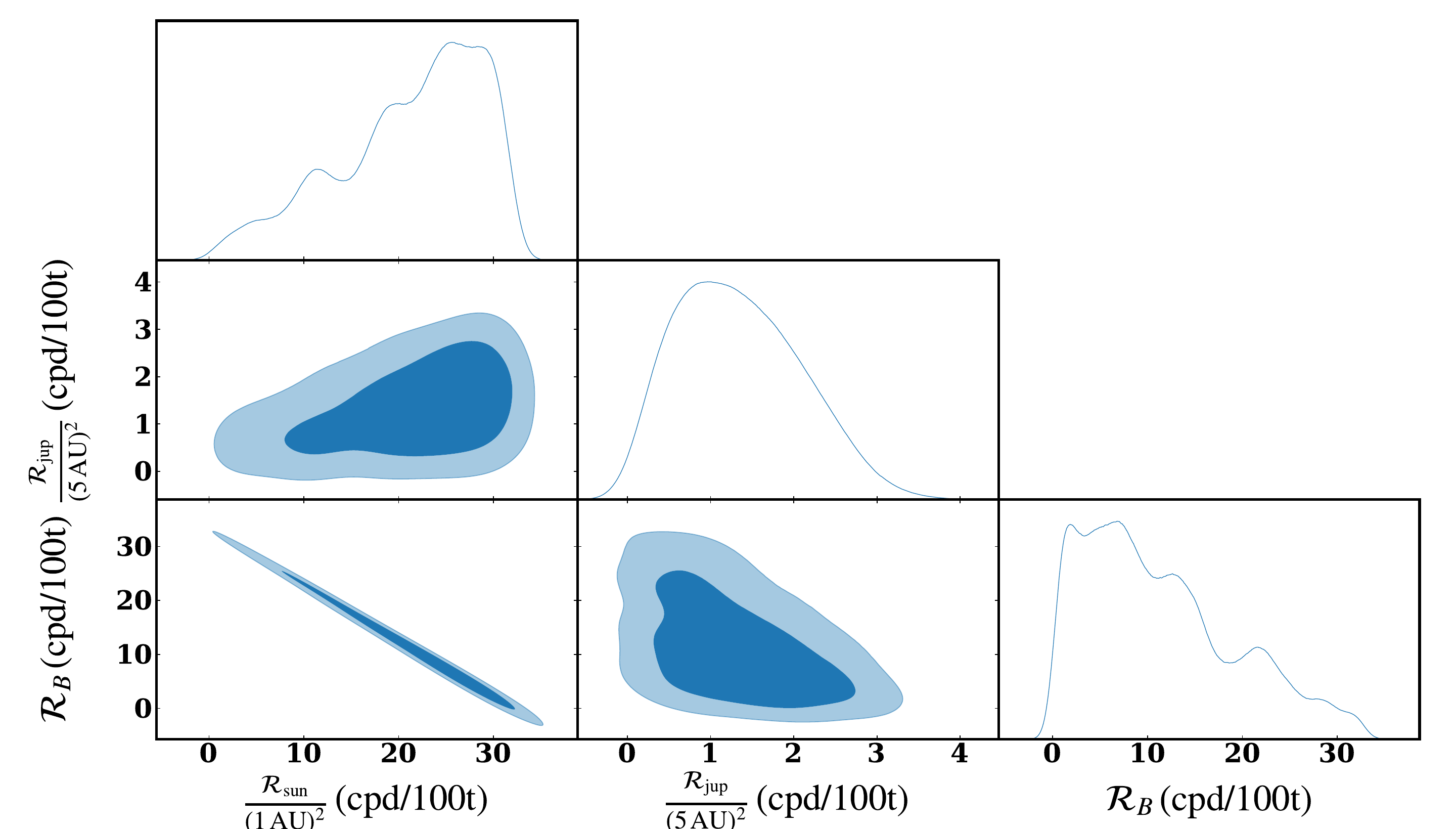}
    \caption{\label{fig:jup2}
   Marginalized 68\% and 95\% credible intervals for $\mathcal{R}_{\mathrm{sun}}$, $\mathcal{R}_{\mathrm{jup}}$ and $\mathcal{R}_{\mathrm{B}}$  after using uniform  prior on $\frac{\mathcal{R}_{\mathrm{jup}}}{(5\,\mathrm{AU})^2} \in \mathcal{U} [0, 4]$ and  uniform prior on $\mathcal{R}_{\mathrm{sun}} \in \mathcal{U} [0, 50]$ with units of (cpd/100t). For this plot, we have used DE442s ephemeris. The marginalized 68\% value for $\frac{\mathcal{R}_{\mathrm{jup}}}{(5\,\mathrm{AU})^2}$ is given by $ 1.31^{+0.84}_{-0.77}$ (cpd/100t).}
\end{figure}

We now repeat the analysis in A24. We download the BOREXINO total rate time-series data from 1 October 2019 to 1 October 2021 and fit for $\mathcal{R}_{jup}$  to Eq.~\ref{eq:R1} using Bayesian regression. To calculate the real-time distance to the Sun and Jupiter, we used the {\tt astropy} library~\cite{astropy}. 
We calculated the aforementioned distances using the JPL ephemerides. We tried multiple ephemerides available in {\tt astropy}, namely DE442s, DE442, DE440s, DE440, DE438, DE435, DE432s and DE430. 
For our Bayesian inference, we used a Gaussian likelihood and the same priors as those used in A24, including the two sets of priors for $\mathcal{R}_{\mathrm{sun}}$.  We sampled the posterior using the {\tt emcee} MCMC sampler~\cite{emcee} and generated the marginalized credible interval plots using {\tt getdist}~\cite{getdist}.
The marginalized 68\% and 90\% credible intervals for the normal and uniform priors on $\mathcal{R}_{\mathrm{sun}}$ can be found in Fig.~\ref{fig:jup1} and Fig.~\ref{fig:jup2}, respectively. 
These best-fit values for \( \frac{\mathcal{R}_{\mathrm{jup}}}{(5\,\mathrm{AU})^2} \) are given by $1.53^{+0.78}_{-0.75}$ and $1.31^{+0.84}_{-0.77}$ (cpd/100t) respectively with significances of $1.8\sigma$ and $1.3\sigma$. 
Note that 5 AU is the average distance to Jupiter between 2019-2021, so the best-fit estimates  for the signal from Jupiter are  normalized with respect to 5 AU, similar to A24.
These best-fit values are  in agreement with the values estimated in A24. We also checked the robustness of these results using all available ephemerides in {\tt astropy}. A tabular summary of these results can be found in Table~\ref{table1}. The results are consistent across all ephemerides used. Therefore, similar to A24 we find approximately $2\sigma$ evidence for a 6\% flux contribution from Jupiter to the BOREXINO rate time-series data.

\begin{table}[!ht]
\begin{tabular}{|c|c|c|}
\hline
\textbf{Ephemeris} & \textbf{Gaussian prior} & \textbf{Uniform prior} \\
& \textbf {(cpd/100t)} & \textbf{(cpd/100t)} \\
\hline
DE442s & \( 1.53^{+0.78}_{-0.75} \) & \( 1.31^{+0.84}_{-0.77} \) \\
DE442  & \( 1.53^{+0.79}_{-0.75} \) & \( 1.31^{+0.84}_{-0.77} \) \\
DE440s & \( 1.54^{+0.79}_{-0.76} \) & \( 1.31^{+0.85}_{-0.77} \) \\
DE440  & \( 1.53^{+0.79}_{-0.75} \) & \( 1.31^{+0.85}_{-0.78} \) \\
DE438  & \( 1.53^{+0.78}_{-0.75} \) & \( 1.29^{+0.84}_{-0.76} \) \\
DE435  & \( 1.53^{+0.79}_{-0.75} \) & \( 1.30^{+0.83}_{-0.76} \) \\
DE432s & \( 1.52^{+0.79}_{-0.75} \) & \( 1.31^{+0.84}_{-0.77} \) \\
DE430  & \( 1.53^{+0.77}_{-0.75} \) & \( 1.31^{+0.85}_{-0.77} \) \\
\hline
\end{tabular}
\caption{\label{table1} Best-fit marginalized constraints on \( \frac{\mathcal{R}_{\mathrm{jup}}}{(5\,\mathrm{AU})^2} \) in units of cpd/100t for two different priors on $\mathcal{R}_{\mathrm{sun}}$.} 
\end{table}

\begin{table}[!ht]
\begin{tabular}{|c|c|c|}
\hline
\textbf{Ephemeris} & \textbf{Gaussian prior} & \textbf{Uniform prior} \\
\hline
DE442s & 3.0  & 1.2  \\
DE442  & 3.0  & 1.0  \\
DE440s & 3.2  & 1.2 \\
DE440  & 3.2 & 1.3 \\
DE438  & 3.2 & 1.3 \\
DE435  & 2.8 & 1.1 \\
DE432s & 2.7 & 1.1 \\
DE430  & 3.0 & 1.2 \\
\hline
\end{tabular}
\caption{\label{table2} Bayes factor for the hypothesis that the BOREXINO flux contains a contribution from Jupiter compared to no contribution for various ephemerides used for two different priors on $\mathcal{R}_{\mathrm{sun}}$ and a uniform prior on $\frac{\mathcal{R}_{\mathrm{jup}}}{(5\,\mathrm{AU})^2} \in \mathcal{U} [0, 4]$. We find that Bayesian model comparison points to only marginal support for the contribution from Jupiter to BOREXINO rate time series data.}
\end{table}

\subsection{Results of Bayesian model comparison}
We now implement Bayesian model comparison to ascertain the significance of the hypothesis that the BOREXINO data is a combination of flux from Sun and Jupiter, compared to the flux coming from only Sun. For this purpose, we provide a brief primer on Bayesian model comparison,  while more details can be found in various reviews~\cite{Trotta,Weller,Sanjib,Krishak}.
To evaluate the significance of a model ($M_2$) as compared to another model ($M_1$), we define  the Bayes factor ($B_{21}$) given by:
\begin{equation}
B_{21}=    \frac{\int P(D|M_2, \theta_2)P(\theta_2|M_2) \, d\theta_2}{\int P(D|M_1, \theta_1)P(\theta_1|M_1) \, d\theta_1} ,  \label{eq:BF}
\end{equation}
where $P(D|M_2,\theta_2)$ is the likelihood for the model $M_2$ for the data $D$, and $P(\theta_2|M_2)$ denotes the prior on the parameter vector $\theta_2$ of the model $M_2$.
The denominator in Eq.~\ref{eq:BF} denotes the same for model $M_1$. If $B_{21}$ is greater than one, then the model $M_2$ is preferred over $M_1$ and vice-versa. The significance can be qualitatively assessed using Jeffreys' scale~\cite{Trotta}. 

For our analysis, the model $M_2$ corresponds to the hypothesis that the BOREXINO rate time-series data are a superposition of flux from Jupiter and Sun, in addition to the background from radioactivity, while $M_1$ corresponds to the hypothesis that the rate time-series
data is due to a combination of only solar neutrino flux and radioactive background. For this analysis, similar to Bayesian inference, we use a Gaussian likelihood and the same priors as A24. To calculate the Bayesian evdience and Bayes factors, we use the {\tt Dynesty} nested sampler~\cite{dynesty}. The results for these Bayes factors can be found in Table~\ref{table2}. We find that for all ephemerides,  the Bayes factor is $\sim 3$ for a Gaussian prior on $\mathcal{R}_{sun}$,  and $\sim 1$ for a uniform prior on $\mathcal{R}_{sun}$. According to Jeffreys' scale, this value corresponds to ``barely worth mentioning''~\cite{Trotta}, implying that the evidence for the extra contribution from Jupiter is marginal. Therefore, Bayesian model comparison provides inconclusive evidence for a contribution from Jupiter in the BOREXINO time series data, compared to a model with no such contribution.

\vspace{1cm}

\vspace{1cm}

\subsection{Search for a signal   from other planets}
We now carry out a similar exercise as in the previous section for other nearby solar system planets, in order to compare and contrast our results with  Jupiter.  This would also serve as a null test of the Bayesian regression analysis technique used in A24, since apriori we do not expect any signal from any other planets,  if an additional signal is argued to come from only Jupiter. For this purpose, we repeat the Bayesian regression and model comparison analysis for Venus, Mars,  and Saturn using the time period between October 2019 and October 2021. We did not consider Mercury since it is very close to the Sun and any signal would be degenerate compared to the Sun.
For Bayesian regression, we replaced the term for Jupiter in Eq.~\ref{eq:R1} with the corresponding term for  a given planet.  Since the average distance to Venus, Mars, and Saturn between 2019-2021 is equal to 1.0, 2.0, and 10.0 AU, respectively, we normalize the signal contributions by the square of these distances similar to that done for Jupiter.  
We use the same priors for $\mathcal{R}_{\mathrm{sun}}$ as those used for Jupiter. However, we changed the lower limit on $\mathcal{R}_{\mathrm{mar}}$ (Mars), $\mathcal{R}_{\mathrm{ven}}$ (Venus), and $\mathcal{R}_{\mathrm{sat}}$ (Saturn) to values less than zero, in order to see if the results are consistent with zero flux. We have considered only one ephemeris, namely DE442s. We have also changed the lower limits on the prior for each of the planets to extend to negative values, in order to check if the signal is compatible with zero value.

\begin{figure}[htbp]
    \centering
    \includegraphics[width=0.6\textwidth]{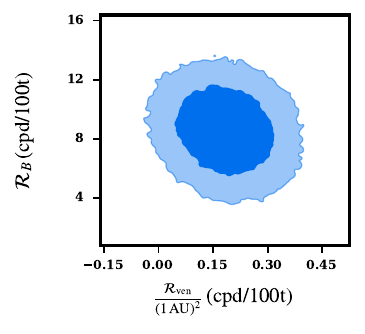}
    \caption{\label{fig:ven1}
   Marginalized 68\% and 95\% credible intervals for   $\mathcal{R}_{\mathrm{B}}$ and  $\mathcal{R}_{\mathrm{ven}}$ after using uniform  prior on $\frac{\mathcal{R}_{\mathrm{ven}}}{(1\,\mathrm{AU})^2} \in \mathcal{U} [-50, 50]$   and   normal prior on $\mathcal{R}_{\mathrm{sun}} \in \mathcal{N} (25,2)$ with units of (cpd/100t). For this plot, we have used DE442s ephemeris. The marginalized 68\% value for $\frac{\mathcal{R}_{\mathrm{ven}}}{(1\,\mathrm{AU})^2}$ is given by \( 0.18^{+0.09}_{-0.09} \) (cpd/100t). }
\end{figure}

\begin{figure}[htbp]
    \centering
    \includegraphics[width=0.9\textwidth]{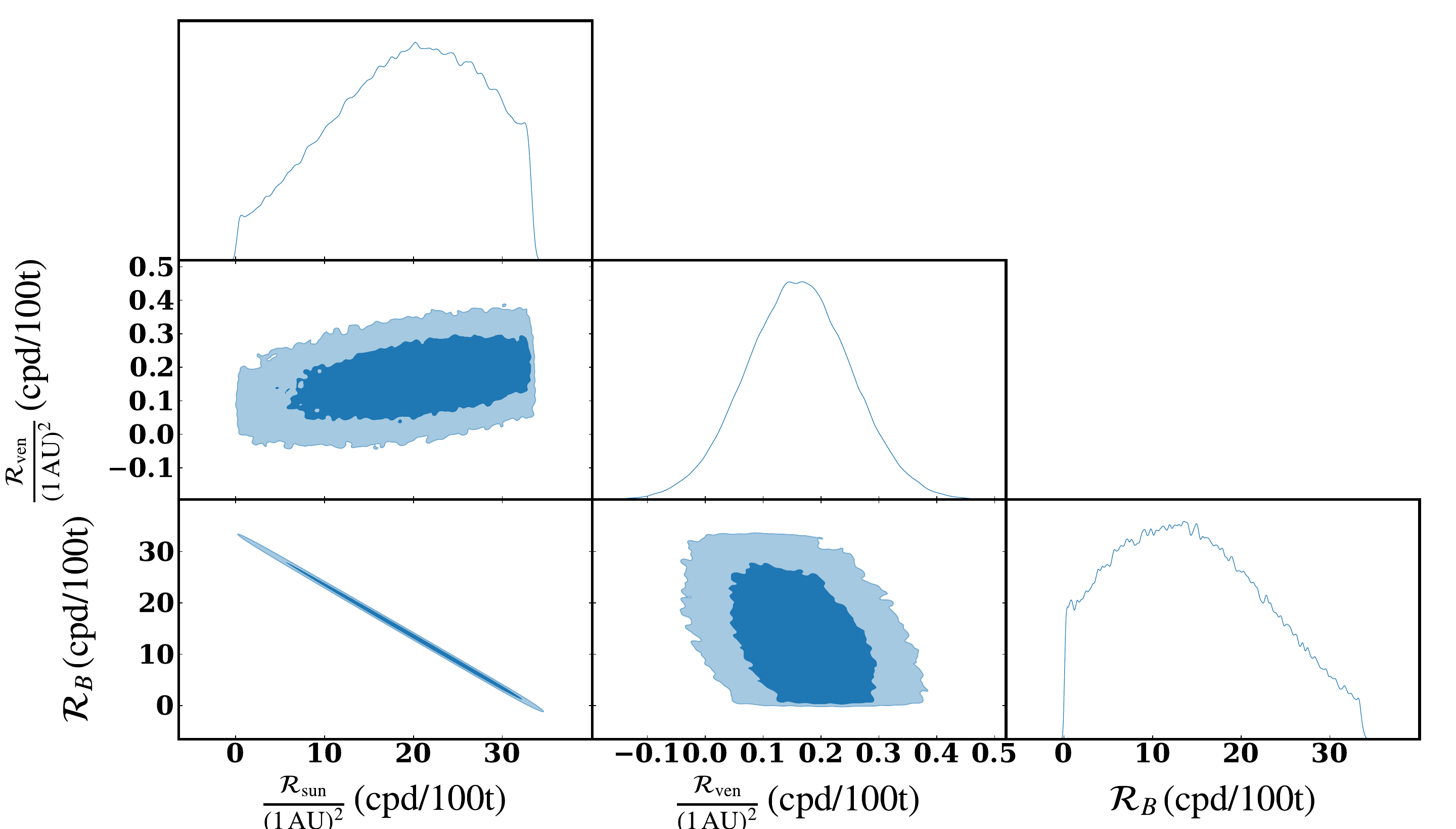}
    \caption{\label{fig:ven2}
  Marginalized 68\% and 95\% credible intervals for $\mathcal{R}_{\mathrm{sun}}$, $\mathcal{R}_{\mathrm{ven}}$ and $\mathcal{R}_{\mathrm{B}}$   after using uniform  prior on $\frac{\mathcal{R}_{\mathrm{ven}}}{(1\,\mathrm{AU})^2} \in \mathcal{U} [-50, 50]$ and   uniform prior on $\mathcal{R}_{sun} \in \mathcal{U} [0, 50]$ with units of (cpd/100t). For this plot, we have used DE442s ephemeris. The marginalized 68\% value for $\mathcal{R}_{ven}\frac{\mathcal{R}_{\mathrm{ven}}}{(1\,\mathrm{AU})^2}$ is given by \( 0.16^{+0.09}_{-0.09} \) (cpd/100t).}
\end{figure}

\begin{figure}[htbp]
    \centering
    \includegraphics[width=0.6\textwidth]{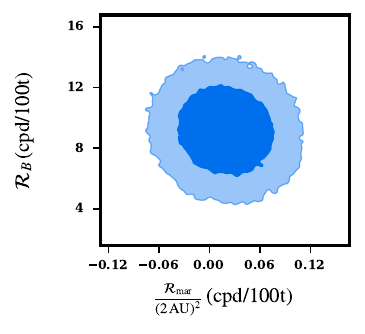}
    \caption{\label{fig:mars1}
   Marginalized 68\% and 95\% credible intervals for   $\mathcal{R}_{\mathrm{B}}$ and  $\mathcal{R}_{\mathrm{mar}}$ after using uniform  prior on $\frac{\mathcal{R}_{\mathrm{mar}}}{(2\,\mathrm{AU})^2} \in \mathcal{U} [-13, 13]$   and   normal prior on $\mathcal{R}_{\mathrm{sun}} \in \mathcal{N} (25,2)$ with units of (cpd/100t). For this plot, we have used DE442s ephemeris. The marginalized 68\% value for $\frac{\mathcal{R}_{\mathrm{mar}}}{(2\,\mathrm{AU})^2}$ is given by \( 0.02^{+0.04}_{-0.04} \) (cpd/100t). }
\end{figure}

\begin{figure}[htbp]
    \centering
    \includegraphics[width=0.9\textwidth]{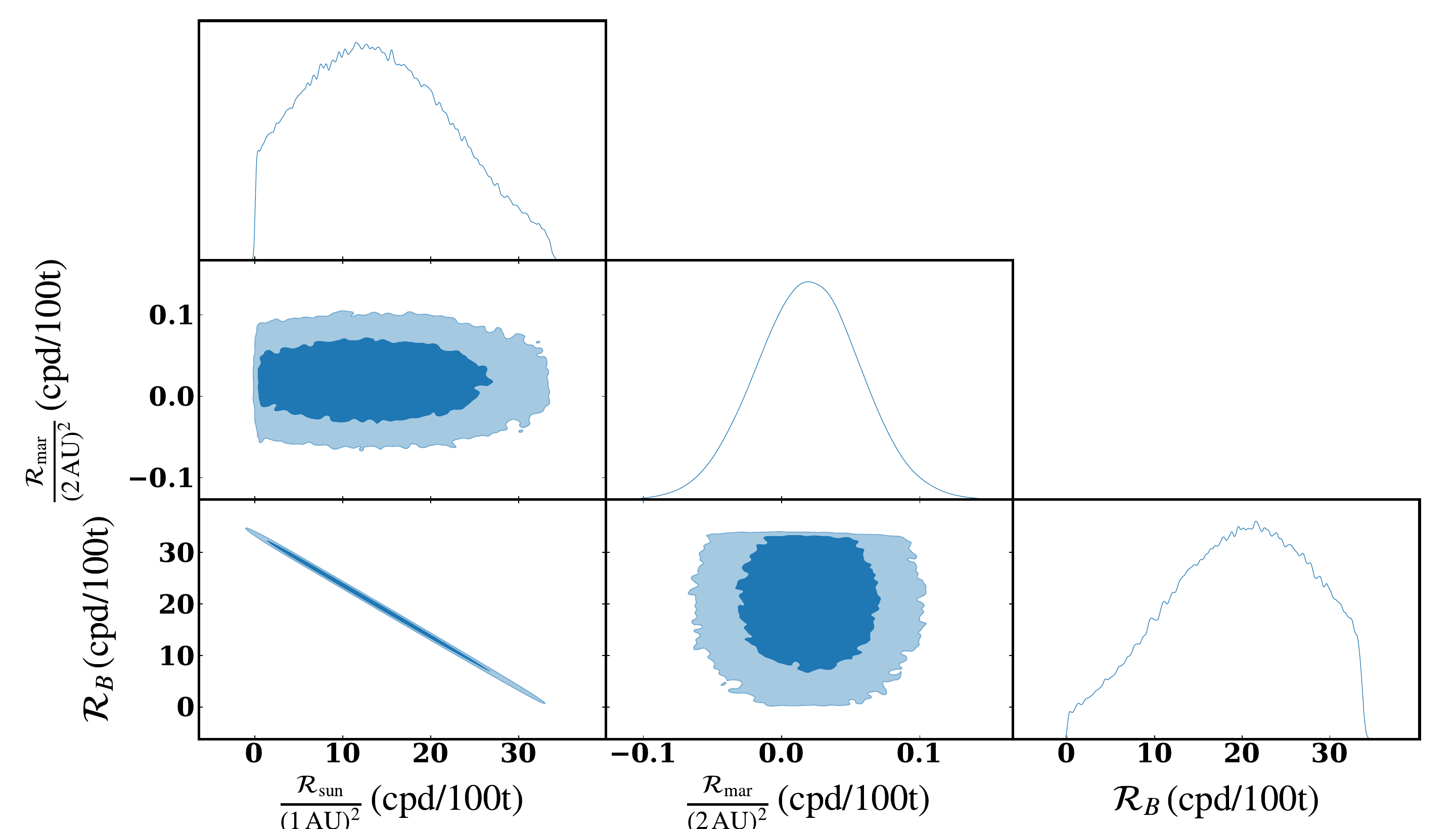}
    \caption{\label{fig:mars2}
   Marginalized 68\% and 95\% credible intervals for $\mathcal{R}_{\mathrm{sun}}$, $\mathcal{R}_{\mathrm{mar}}$ and $\mathcal{R}_{\mathrm{B}}$   after using a uniform  prior on $\frac{\mathcal{R}_{\mathrm{mar}}}{(2\,\mathrm{AU})^2} \in \mathcal{U} [-13, 13]$ and uniform prior on $\mathcal{R}_{\mathrm{sun}} \in \mathcal{U} [0, 50]$ with units of (cpd/100t). For this plot, we have used DE442s ephemeris. The marginalized 68\% value for $\frac{\mathcal{R}_{\mathrm{mar}}}{(2\,\mathrm{AU})^2}$ is given by \( 0.02^{+0.04}_{-0.04} \) (cpd/100t).}
\end{figure}

\begin{figure}[htbp]
    \centering
    \includegraphics[width=0.6\textwidth]{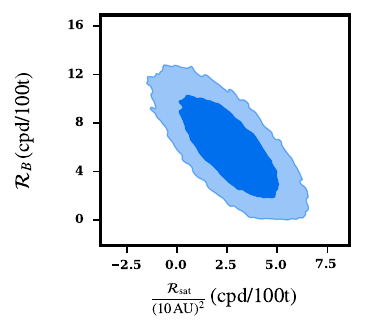}
    \caption{\label{fig:saturn1}
   Marginalized 68\% and 95\% credible intervals for $\mathcal{R}_{\mathrm{B}}$ and  $\mathcal{R}_{\mathrm{sat}}$ after using uniform  prior on $\frac{\mathcal{R}_{\mathrm{sat}}}{(10\,\mathrm{AU})^2} \in \mathcal{U} [-10, 10]$   and   normal prior on $\mathcal{R}_{\mathrm{sun}} \in \mathcal{N} (25,2)$ with units of (cpd/100t). For this plot, we have used DE442s ephemeris. The marginalized 68\% value for $\frac{\mathcal{R}_{\mathrm{sat}}}{(10\,\mathrm{AU})^2}$ is given by \( 2.59^{+1.59}_{-1.62} \) (cpd/100t). }
\end{figure}

\begin{figure}[htbp]
    \centering
    \includegraphics[width=0.9\textwidth]{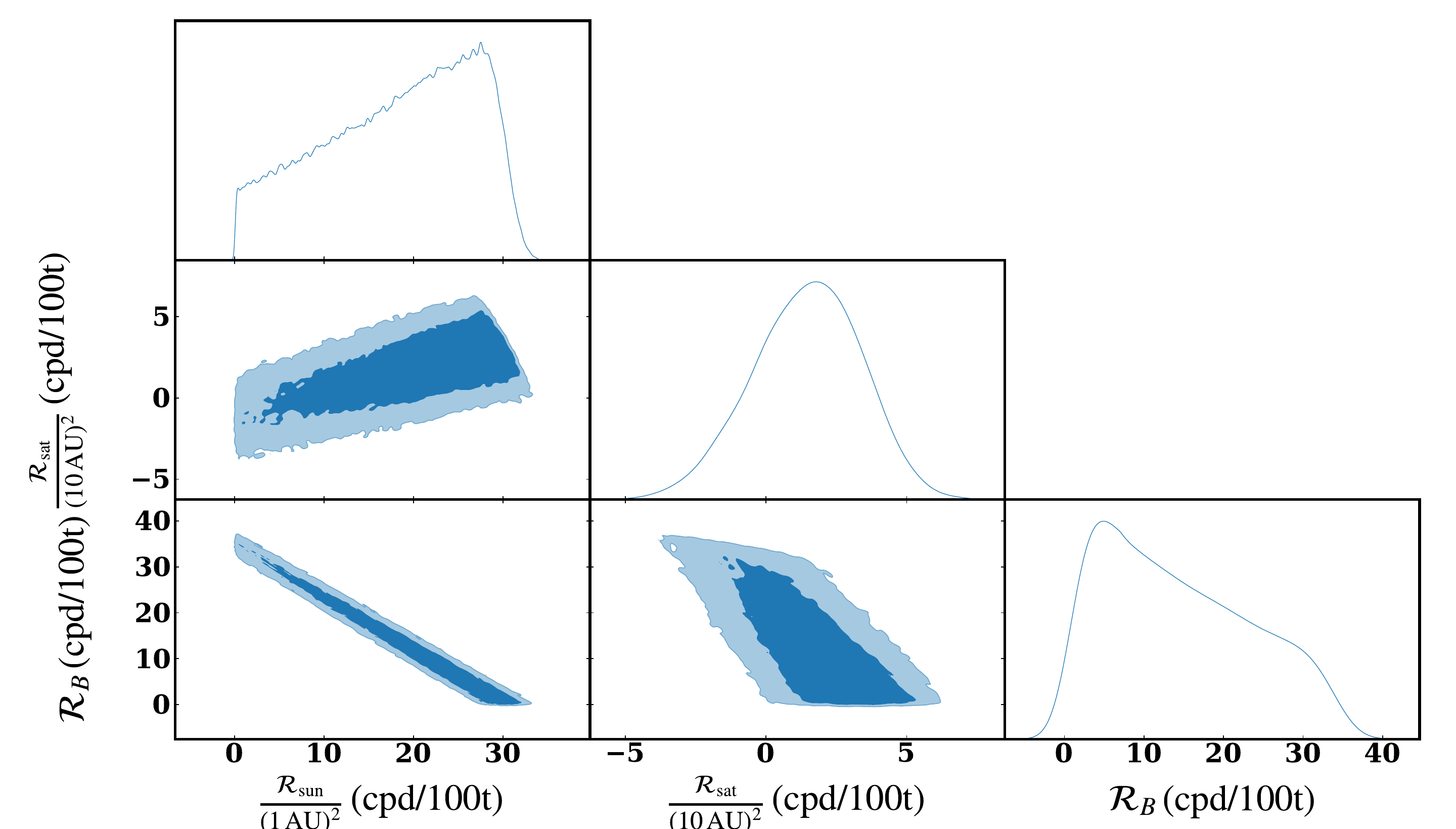}
    \caption{\label{fig:saturn2}
   Marginalized 68\% and 95\% credible intervals for $\mathcal{R}_{\mathrm{sun}}$, $\mathcal{R}_{\mathrm{sat}}$ and $\mathcal{R}_{\mathrm{B}}$   after using uniform  prior on $\frac{\mathcal{R}_{\mathrm{sat}}}{(10\,\mathrm{AU})^2} \in \mathcal{U} [-10, 10]$ and   uniform prior on $\mathcal{R}_{\mathrm{sun}} \in \mathcal{U} (0,50)$ with units of (cpd/100t). For this plot, we have used DE442s ephemeris. The marginalized 68\% value for $\frac{\mathcal{R}_{\mathrm{sat}}}{(10\,\mathrm{AU})^2}$ is given by \( 1.53^{+1.87}_{-2.06} \) (cpd/100t).}
\end{figure}

The corresponding marginalized 68\% and 90\% credible intervals  can be found in Figs.~\ref{fig:ven1} and ~\ref{fig:ven2}, Figs.~\ref{fig:mars1} and ~\ref{fig:mars2}, and Figs.~\ref{fig:saturn1} and ~\ref{fig:saturn2} respectively for Venus, Mars, and Saturn respectively. A tabular summary can be found in Table~\ref{tableplanet}. For Mars, the $1\sigma$ central intervals are consistent with zero. For Saturn, we get  $1.6\sigma$ evidence for non-zero flux  for a Gaussian prior on $\mathcal{R}_{\mathrm{sun}}$ with values of  $\frac{\mathcal{R}_{\mathrm{sat}}}{(10\,\mathrm{AU})^2} = 2.59^{+1.59}_{-1.62}$ (cpd/100t). However, for a uniform prior on  $\mathcal{R}_{\mathrm{sun}}$, the contribution from Saturn is consistent with zero.
For Venus, we get central estimates for $\frac{\mathcal{R}_{\mathrm{ven}}}{(1\,\mathrm{AU})^2}$ equal to $0.18^{+0.09}_{-0.09}$ and $0.16^{+0.09}_{-0.09}$ (cpd/100t), for Gaussian and uniform priors on $\mathcal{R}_{\mathrm{sun}}$, corresponding to about $2\sigma$ significances, similar to that seen for Jupiter. The inferred flux from Venus about 10\% smaller than Jupiter and about 0.6\% compared to the solar contribution. Therefore, the Bayesian regression technique implemented in A24 also returns spurious signals for Venus ($2\sigma$) and Saturn ($1.6\sigma$) (for one of the prior choices).

We also carried out Bayesian model comparison for these planets  similar to that done for Jupiter. Similar to before, the null hypothesis involves the BOREXINO rate time-series data containing a contribution from only Sun in addition to the radioactive  backgrounds, whereas the alternative hypothesis involves an additional contribution  from the corresponding planet.
Unlike Bayesian regression, we used uniform priors with only a positive contribution from a given planet, with the minimum value of 0, since we only want to compare the significance of a positive signal. The results for the Bayes factors can be found in Table~\ref{table4}, along with the priors used.
We find that the Bayes factors are less than one for Venus and Mars with values $\lessapprox 0.01$, implying that additional contributions from Venus and Mars are decisively ruled out. For Saturn we find that the Bayes factors are close to 1, implying that the significance of both the  hypotheses are comparable and there is no support for additional contribution from Saturn.

Therefore, Bayesian regression also provides a $2\sigma$ evidence for a (spurious) contribution from Venus and $1.6\sigma$ contribution from Saturn (for a Gaussian prior on Sun)  similar to that from Jupiter.
However, a Bayesian model comparison does not support any additional contribution to the total rate from Mars, Venus, and Saturn.



\begin{table}[h]
\begin{tabular}{|c|c|c|c|c|}
\hline
\textbf{Planet} & \textbf{Term} & \textbf{Prior} & \textbf{Gaussian prior} & \textbf{Uniform prior} \\
&&& \textbf {(cpd/100t)} & \textbf{(cpd/100t)} \\
\hline
Venus  & \( \frac{\mathcal{R}_{\mathrm{ven}}}{(1\,\mathrm{AU})^2} \)  & \(\mathcal{U}[-50, 50]\) & \( 0.18^{+0.09}_{-0.09} \) & \( 0.16^{+0.09}_{-0.09} \) \\
Mars   & \( \frac{\mathcal{R}_{\mathrm{mar}}}{(2\,\mathrm{AU})^2} \)  & \(\mathcal{U}[-13, 13]\) & \( 0.02^{+0.04}_{-0.04} \) & \( 0.02^{+0.04}_{-0.04} \) \\
Saturn & \( \frac{\mathcal{R}_{\mathrm{sat}}}{(10\,\mathrm{AU})^2} \)  & \(\mathcal{U}[-10, 10]\) & \( 2.59^{+1.59}_{-1.62} \) & \( 1.53^{+1.87}_{-2.06} \) \\
\hline
\end{tabular}
\caption{\label{tableplanet} Best-fit marginalized constraints on \( \frac{\mathcal{R}_{\mathrm{ven}}}{(1\,\mathrm{AU})^2} \), \( \frac{\mathcal{R}_{\mathrm{mar}}}{(2\,\mathrm{AU})^2} \), and \( \frac{\mathcal{R}_{\mathrm{sat}}}{(10\,\mathrm{AU})^2} \) in units of cpd/100t for two different priors on $\mathcal{R}_{\mathrm{sun}}$ using DE442s ephemeris.}
\end{table}

\begin{table}[h]
\begin{tabular}{|c|c|c|c|}
\hline
\textbf{Planet} & \textbf{Prior} & \textbf{Gaussian prior} & \textbf{Uniform prior} \\
\hline
Venus & \( \frac{\mathcal{R}_{\mathrm{ven}}}{(1\,\mathrm{AU})^2} \in \mathcal{U}[0, 50]\) & 0.0397  & 0.0207  \\
Mars   & \( \frac{\mathcal{R}_{\mathrm{mar}}}{(2\,\mathrm{AU})^2} \in \mathcal{U}[0, 13]\) & 0.0062  & 0.0059  \\
Saturn & \( \frac{\mathcal{R}_{\mathrm{sat}}}{(10\,\mathrm{AU})^2} \in \mathcal{U}[0, 10]\) & 1.43  & 0.55 \\
\hline
\end{tabular}
\caption{\label{table4} Bayes factor for Venus, Mars, and Saturn using DE442s ephemeris for two different priors on $\mathcal{R}_{\mathrm{sun}}$. Here, the null hypothesis  corresponds to the flux consisting of only contribution from the Sun and the radioactive backgrounds, whereas the alternative  hypothesis involves an additional contribution from the corresponding planet.}
\end{table}

\subsection{Goodness of fit tests}
\begin{figure}[htbp]
    \centering
    \includegraphics[width=0.9\textwidth]{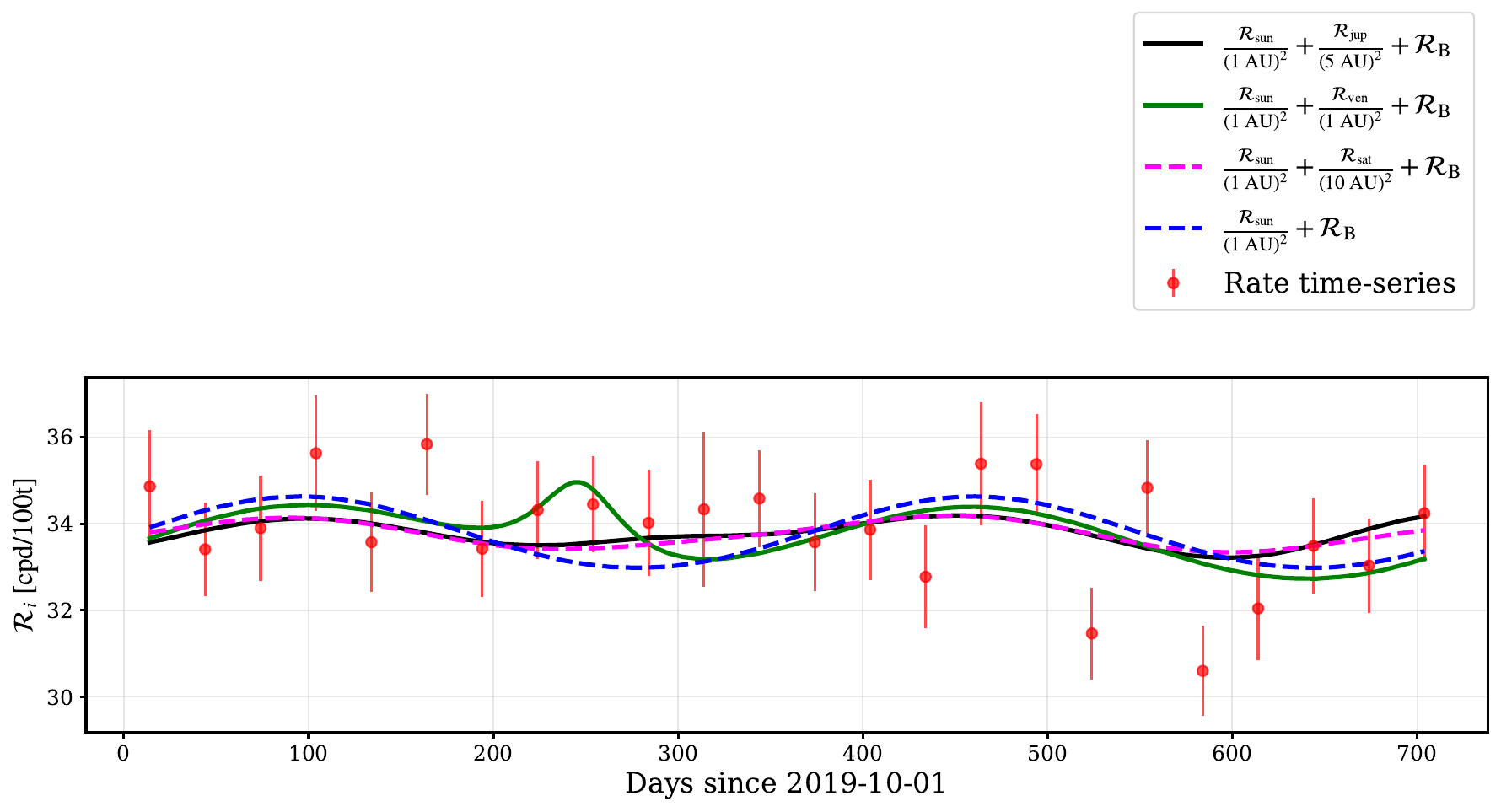}
    \caption{\label{fig:ratechi}
    Monthly-binned $\beta$-like event rate in the energy range corresponding to the $^7$Be solar neutrinos from October 2019 to October 2021 obtained by the BOREXINO collaboration. The black line shows the prediction with event rate contribution from Jupiter, the green line shows the prediction with event rate contribution from Venus, the magenta dashed line shows the prediction with event rate contribution from Saturn, and the blue dashed line shows the prediction with event rate contribution from only Sun using DE442s ephemeris
    $\frac{\chi^{2}_{sun+jup}}{dof}=\frac{25.3}{21},$
    $\frac{\chi^{2}_{sun+ven}}{dof}=\frac{24.3}{21},$
    $\frac{\chi^{2}_{sun+sat}}{dof}=\frac{26.2}{21},$
    $\frac{\chi^{2}_{sun}}{dof}=\frac{28.6}{22}$}
\end{figure}

Therefore, we find that Bayesian regression points to $\sim 2\sigma$ evidence for additional flux from Jupiter, Venus, and Saturn (for one chosen prior), whereas, Bayesian model comparison does not support any additional contribution from Jupiter, Venus or Saturn. We then overlay the total time series data to all three models considered so far, involving a contribution from only Sun, Sun+Jupiter, Sun+Venus, and Sun+Saturn This plot can be found in Fig.~\ref{fig:ratechi}. We also calculate the reduced $\chi^2$ for all four hypotheses, to check if each of these provides a good fit. We find $\chi^2/dof$ equal to  28.6/22 (1.3), 24.3/21 (1.15), 25.3/21 (1.2), 26.2/21 (1.2)  for Sun, Sun+Venus, Sun+Jupiter, Sun+Saturn respectively. Therefore, the reduced $\chi^2$ are close to one   for all the four hypotheses. 

\section{Analysis with monthly modulation data}
\label{sec:modulation}
\begin{figure}[htbp]
    \centering
    \includegraphics[width=0.5\textwidth]{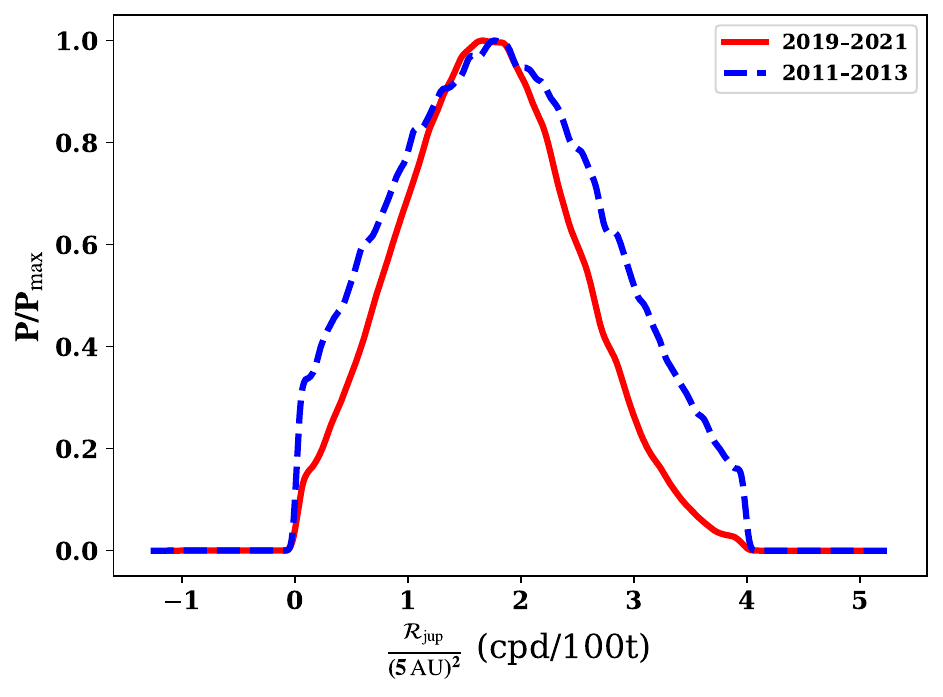}
    \caption{\label{fig:Modulation}
   The marginalized posteriors on $\mathcal{R}_{\mathrm{jup}}$ for Jupiter from October 2019 to October 2021 (Red line) and December 2011 to December 2013 (Blue line) with a uniform prior on $\frac{\mathcal{R}_{\mathrm{jup}}}{(5\,\mathrm{AU})^2} \in \mathcal{U} [0, 4]$ using the monthly modulation binned data (cf. Eq.~\ref{eq:4}) and DE442s ephemeris.}
\end{figure}

We now do a similar analysis with the monthly modulation data ($\delta  \mathcal{R}(t)$) obtained from Eq.~\ref{eq:R2} by subtracting the trend of the  data ($R_{tr}$) from the rate time-series data. The value of $R_{tr}$ in each time bin has been made available by the BOREXINO collaboration. Similar to A24, we used the monthly modulation data from two time periods, 11 December 2011 - 11 December 2013 and 1st October 2019 - 1st October 2021. We fit ($\delta  \mathcal{R}(t)$) to  $\mathcal{R}_{\mathrm{sun}}$ and $\mathcal{R}_{\mathrm{jup}}$. We used a Gaussian likelihood, while normal and uniform priors were used for $\mathcal{R}_{\mathrm{sun}}$ and $\mathcal{R}_{\mathrm{jup}}$ respectively, given by $\mathcal{R}_{\mathrm{sun}} \in \mathcal{N} (25,2)$ and $\mathcal{R}_{\mathrm{jup}} \in \mathcal{U} [0, 50] $ (cpd/100t), respectively. 
 Since our results obtained by fitting to Eq.~\ref{eq:3} were not compatible to those in A24, we used an augmented version of Eq.~\ref{eq:3}, given by~\footnote{This holds because the numerical integration uses uniform 30-day intervals; thus, dividing the integral term by the total time period is equivalent to computing the average value of the second term.}:
\begin{equation}
\delta \mathcal{R}(t) = \sum_i  \mathcal{R}_{i} \left[\frac{1}{d^2_{i}(t)} -    \biggl< \frac{dt}{d^2_{i}(t)} \biggr> \right],
\label{eq:4}
\end{equation}
where the index $i$ once again runs over Sun and Jupiter.
The best-fit marginalized values for $\mathcal{R}_{\mathrm{jup}}$ for the two time intervals can be found in Fig.~\ref{fig:Modulation} for the DE422s ephemerides. These are in agreement with Fig.~3 of A24. The best-fit values for $\mathcal{R}_{\mathrm{jup}}$  are equal to $1.78^{+1.03}_{-0.97}$ and $1.69^{+0.79}_{-0.77}$ (cpd/100t), for 2011-2013 and 2019-2021 respectively, corresponding to significances of $1.8\sigma$ and $2.1\sigma$.  We also did a similar exercise using other ephemerides. These values can be found in Table~\ref{tablemodbestfit}. We find  that  the results for both  epochs are consistent with DE422s for all the ephemerides used. Therefore, even with the monthly modulation data we find evidence for $2\sigma$ contribution from Jupiter with flux contribution of about 6\% of that seen from the Sun.

We then do Bayesian model comparison in the same way as done in Sect.~\ref{sec:analysis}. We used a Gaussian likelihood and the same prior for $\mathcal{R}_{\mathrm{sun}}$  and $\mathcal{R}_{\mathrm{jup}}$ as used in the Bayesian regression. The null hypothesis corresponds to the ansatz that the modulation data is due to only the Sun, whereas the alternate hypothesis is that the modulation data is a combination of data from Sun and Jupiter. The Bayes factors for all the ephemerides for both epochs can be found in Table~\ref{tablemodbf}. The Bayes factor for 2011-2013 period is around two and hence not significant. Although the Bayes factor for the 2019-2021 period is marginally larger  than with the total rate time-series data with a value of approximately four, it is still $< 5$ and according to Jeffreys scale, still point to ``barely worth mentioning''~\cite{Trotta}. Therefore, a Bayesian model comparison still only provides a marginal support for an additional contribution from Jupiter to the monthly modulation data during the periods 2019-2021 and 2011-2013.

\subsection{Searches from other planets using modulation data}
\begin{figure}[htbp]
    \centering
    \includegraphics[width=0.5\textwidth]{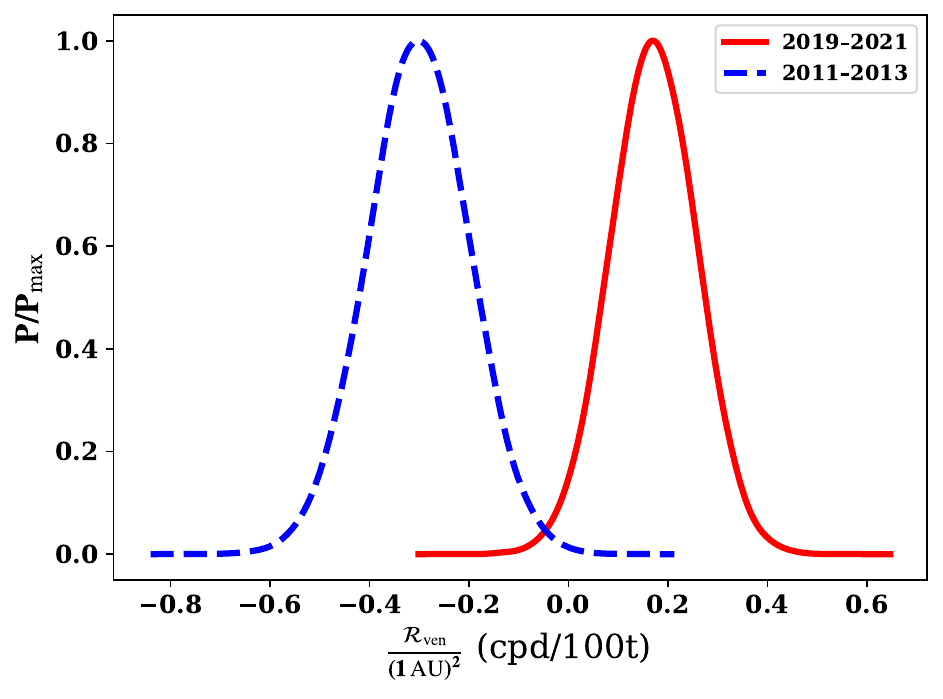}
    \caption{\label{fig:Modulationven}
   The marginalized posteriors on $\mathcal{R}_{\mathrm{ven}}$ for Venus from October 2019 to October 2021 (Red line) and December 2011 to December 2013 (Blue line) with a uniform prior on $\frac{\mathcal{R}_{\mathrm{ven}}}{(1\,\mathrm{AU})^2} \in \mathcal{U}[-50, 50]$ using the monthly modulation binned data (cf. Eq.~\ref{eq:4}) and DE442s ephemeris.}
\end{figure}

\begin{figure}[htbp]
    \centering
    \includegraphics[width=0.5\textwidth]{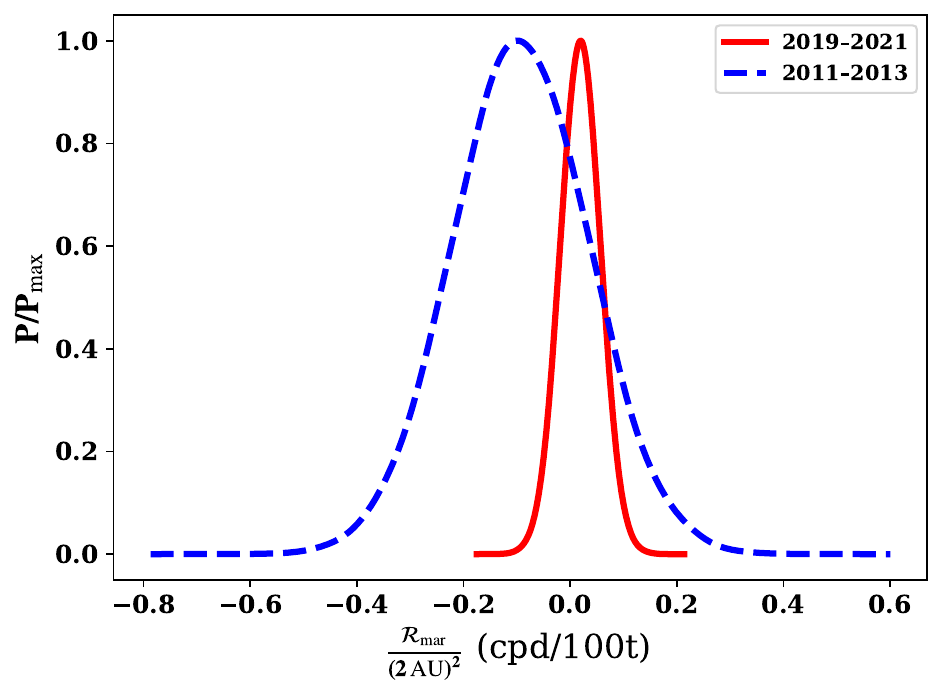}
    \caption{\label{fig:Modulationmars}
   The marginalized posteriors on $\mathcal{R}_{\mathrm{mar}}$ for Mars from October 2019 to October 2021 (Red line) and December 2011 to December 2013 (Blue line) with a uniform prior on $\frac{\mathcal{R}_{\mathrm{mar}}}{(2\,\mathrm{AU})^2} \in \mathcal{U}[-13, 13]$ using the monthly modulation binned data (cf. Eq.~\ref{eq:4}) and DE442s ephemeris.}
\end{figure}

\begin{figure}[htbp]
    \centering
    \includegraphics[width=0.5\textwidth]{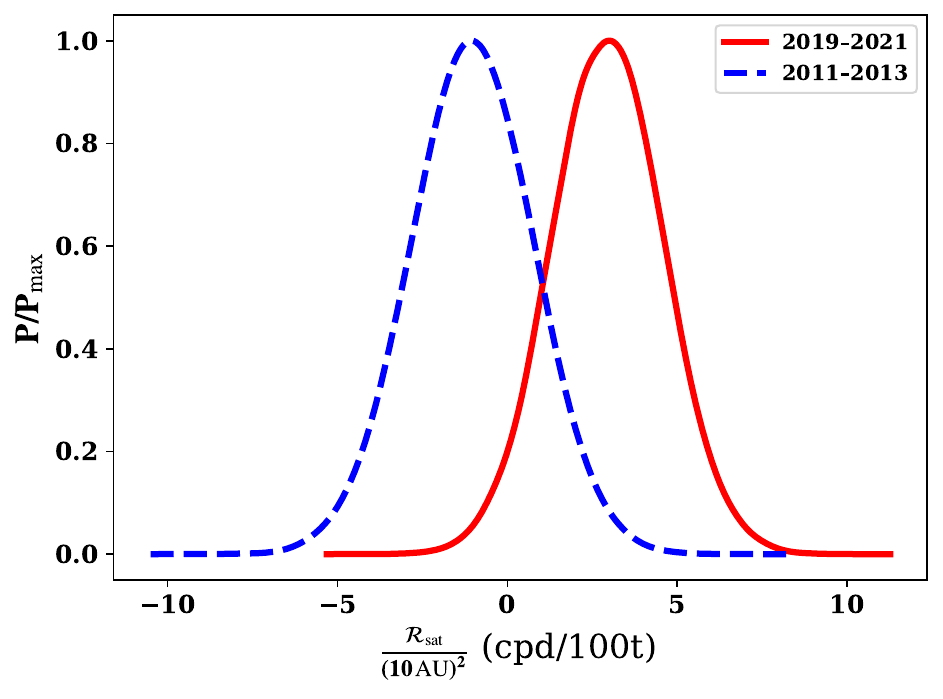}
    \caption{\label{fig:Modulationsat}
   The marginalized posteriors on $\mathcal{R}_{\mathrm{sat}}$ for Saturn from October 2019 to October 2021 (Red line) and December 2011 to December 2013 (Blue line) with a uniform prior on $\frac{\mathcal{R}_{\mathrm{sat}}}{(10\,\mathrm{AU})^2} \in \mathcal{U}[-10, 10]$ using the monthly modulation binned data (cf. Eq.~\ref{eq:4}) and DE442s ephemerides.}
\end{figure}

We now do the same analysis using the monthly modulation data to search for a signal from Venus, Mars, and Saturn, in order to serve as a null test of the analysis technique used for Jupiter. We fit each planet data to Eq.~\ref{eq:4}, by replacing $\mathcal{R}_{\mathrm{jup}}$ with the corresponding contribution from Venus, Mars, and Sun. The prior used for $\mathcal{R}_{\mathrm{sun}}$ is the same as that used for Jupiter. 
 However, for the corresponding prior on each of the planets we extended the lower limits to negative values in order to test for compatibility with zero flux.  For calculating the distance, we used  DE442s ephemerides. 
 The marginalized best fit posterior intervals for
$\mathcal{R}_{\mathrm{ven}}$, $\mathcal{R}_{\mathrm{mar}}$, and $\mathcal{R}_{\mathrm{sat}}$ can be found in Figs.~\ref{fig:Modulationven}, ~\ref{fig:Modulationmars}, and  ~\ref{fig:Modulationsat}, respectively. The corresponding best-fit values are tabulated in Table~\ref{tablemodplanet}.  For  Venus, Mars, and Saturn, we find  the best-fit values for the flux  contribution to be less than 0 during the 2011-2013 interval. During 2019-2021, the best-fit values for Mars are consistent with zero flux. We find that the best-fit values for  Venus flux  in the 2019-2021 interval are  given by $\frac{\mathcal{R}_{\mathrm{ven}}}{(1\,\mathrm{AU})^2} = 0.17 \pm 0.09$ (cpd/100t). Therefore, this is consistent with a non-zero flux at $2\sigma$ significance,  similar to that seen for Jupiter. The inferred flux is comparable to that obtained by using the rate time-series data. For Saturn we find $\frac{\mathcal{R}_{\mathrm{sat}}}{(10\,\mathrm{AU})^2} = 2.98 \pm 1.64$ (cpd/100t) (corresponding to a significance of  $1.8\sigma$).

We now do hypothesis testing using Bayesian model selection. The null hypothesis corresponds to the {\it ansatz} that the monthly modulation data contain a contribution from only Sun, while the alternative hypothesis is that there is an additional contribution from the planet (Venus, Mars, or Saturn). Since we are only interested in a positive flux contribution, we ensure that the lower limits on the prior for $\mathcal{R}_{\mathrm{ven}}$, $\mathcal{R}_{\mathrm{jup}}$, and $\mathcal{R}_{\mathrm{sat}}$ are equal to 0.  The Bayes factors for 2011-2013 and 2019-2021 intervals can be found from Table~\ref{tablemodbayesplanet}. We find that the Bayes factors are much less than one for Mars and Venus  for both 2019-2021 and 2011-2013. For Saturn, the Bayes factors are close to 1 for both 2019-2021 and 2011-2013.  Therefore, Bayesian model comparison does not support an additional contribution from any additional planet.  

\begin{table}[!ht]
\begin{tabular}{|c|c|c|}
\hline
\textbf{Ephemeris} & \textbf{Dec. 2011 - Dec. 2013} & \textbf{Oct. 2019 - Oct. 2021} \\
& \textbf {(cpd/100t)} & \textbf{(cpd/100t)} \\
\hline
DE442s & \( 1.78^{+1.03}_{-0.97} \) & \( 1.69^{+0.79}_{-0.77} \)  \\
DE442  & \( 1.79^{+1.03}_{-0.97} \) & \( 1.69^{+0.81}_{-0.77} \)  \\
DE440s & \( 1.79^{+1.03}_{-0.97} \) & \( 1.70^{+0.79}_{-0.77} \) \\
DE440  & \( 1.79^{+1.02}_{-0.96} \) & \( 1.70^{+0.78}_{-0.77} \) \\
DE438  & \( 1.78^{+1.02}_{-0.97} \) & \( 1.70^{+0.79}_{-0.78} \) \\
DE435  & \( 1.79^{+1.02}_{-0.97} \) & \( 1.70^{+0.79}_{-0.78} \) \\
DE432s & \( 1.79^{+1.03}_{-0.96} \) & \( 1.70^{+0.79}_{-0.77} \) \\
DE430  & \( 1.80^{+1.02}_{-0.97} \) & \( 1.69^{+0.79}_{-0.77} \) \\
\hline
\end{tabular}
\caption{\label{tablemodbestfit} Best fit marginalized values for $\mathcal{R}_{\mathrm{jup}}/ \rm{(5 AU)^2}$ along with $1\sigma$ error bars obtained using a Bayesian regression to monthly modulation data (cf. Eq.~\ref{eq:4}) for different ephemerides. The results using all the ephemerides are consistent with each other. }
\end{table}

\begin{table}[!ht]
\begin{tabular}{|c|c|c|}
\hline
\textbf{Ephemeris} & \textbf{Dec. 2011 - Dec. 2013} & \textbf{Oct. 2019 - Oct. 2021} \\
\hline
DE442s & 2.2  & 4.2  \\
DE442  & 2.1  & 4.2  \\
DE440s & 2.2  & 4.2 \\
DE440  & 2.1 & 4.2 \\
DE438  & 2.2 & 4.2 \\
DE435  & 2.1 & 4.1 \\
DE432s & 2.1 & 4.2 \\
DE430  & 2.2 & 4.2 \\
\hline
\end{tabular}
\caption{\label{tablemodbf} Bayes factor for the hypothesis that the BOREXINO flux contains a contribution from Jupiter compared to no contribution from Jupiter for various ephemerides with a uniform prior on $\frac{\mathcal{R}_{\mathrm{jup}}}{(5\,\mathrm{AU})^2} \in \mathcal{U} [0, 4]$. We find that Bayesian model comparison points to only marginal support for contribution from Jupiter.}
\end{table}

\begin{table}[h]
\begin{tabular}{|c|c|c|c|c|}
\hline
\textbf{Planet} & \textbf{Term} &\textbf{Prior} & \textbf{Dec. 2011 - Dec. 2013} & \textbf{Oct. 2019 - Oct. 2021} \\
&&& \textbf {(cpd/100t)} & \textbf{(cpd/100t)} \\
\hline
Venus & \(\frac{\mathcal{R}_{\mathrm{ven}}}{(1\,\mathrm{AU})^2} \) & $\mathcal{U} [-50, 50]$  & \( -0.30^{+0.10}_{-0.10} \) & \( 0.17^{+0.09}_{-0.09} \) \\
Mars   & \( \frac{\mathcal{R}_{\mathrm{mar}}}{(2\,\mathrm{AU})^2} \)  & $\mathcal{U} [-13, 13]$ & \( -0.09^{+0.13}_{-0.13} \) & \( 0.02^{+0.04}_{-0.04} \) \\
Saturn & \( \frac{\mathcal{R}_{\mathrm{sat}}}{(10\,\mathrm{AU})^2} \) & $\mathcal{U} [-10, 10]$ & \( -1.04^{+1.80}_{-1.81} \) & \( 2.98^{+1.63}_{-1.64} \) \\
\hline
\end{tabular}
\caption{\label{tablemodplanet} Best-fit marginalized values for the flux contribution from Venus, Mars, and Saturn ($\mathcal{R}_{\mathrm{ven}}$, $\mathcal{R}_{\mathrm{mar}}$, and $\mathcal{R}_{\mathrm{sat}}$) to the monthly modulation data obtained from  Eq.~\ref{eq:4} by replacing $\mathcal{R}_{\mathrm{i}}$ with the corresponding planet.}
\end{table}

\begin{table}[h]
\begin{tabular}{|c|c|c|c|}
\hline
\textbf{Planet}  & \textbf{Prior} & \textbf{Dec. 2011 - Dec. 2013} & \textbf{Oct. 2019 - Oct. 2021} \\
\hline
Venus & $ \frac{\mathcal{R}_{\mathrm{ven}}}{(1\,\mathrm{AU})^2} \in \mathcal{U} [0, 50]$ & 0.0007  & 0.03  \\
Mars  & $ \frac{\mathcal{R}_{\mathrm{mar}}}{(2\,\mathrm{AU})^2} \in \mathcal{U} [0, 13]$ &  0.007  & 0.006  \\
Saturn & $ \frac{\mathcal{R}_{\mathrm{sat}}}{(10\,\mathrm{AU})^2} \in \mathcal{U} [0, 10]$ & 0.2 & 2.0 \\
\hline
\end{tabular}
\caption{\label{tablemodbayesplanet} Bayes factors for the {\it ansatz} that the BOREXINO flux contains an additional  contribution from Venus, Mars, or Saturn (in addition to Sun) compared to  no contribution for various ephemerides using a positive prior for each of them. We find that Bayesian model comparison points to only marginal support for additional contribution from Saturn (2019-2021) but not from 2011-2013, while it is disfavored for Mars and Venus for both the epochs.}
\end{table}

\subsection{Goodness of fit tests with monthly modulation data}
\begin{figure}[htbp]
    \centering
    \includegraphics[width=0.9\textwidth]{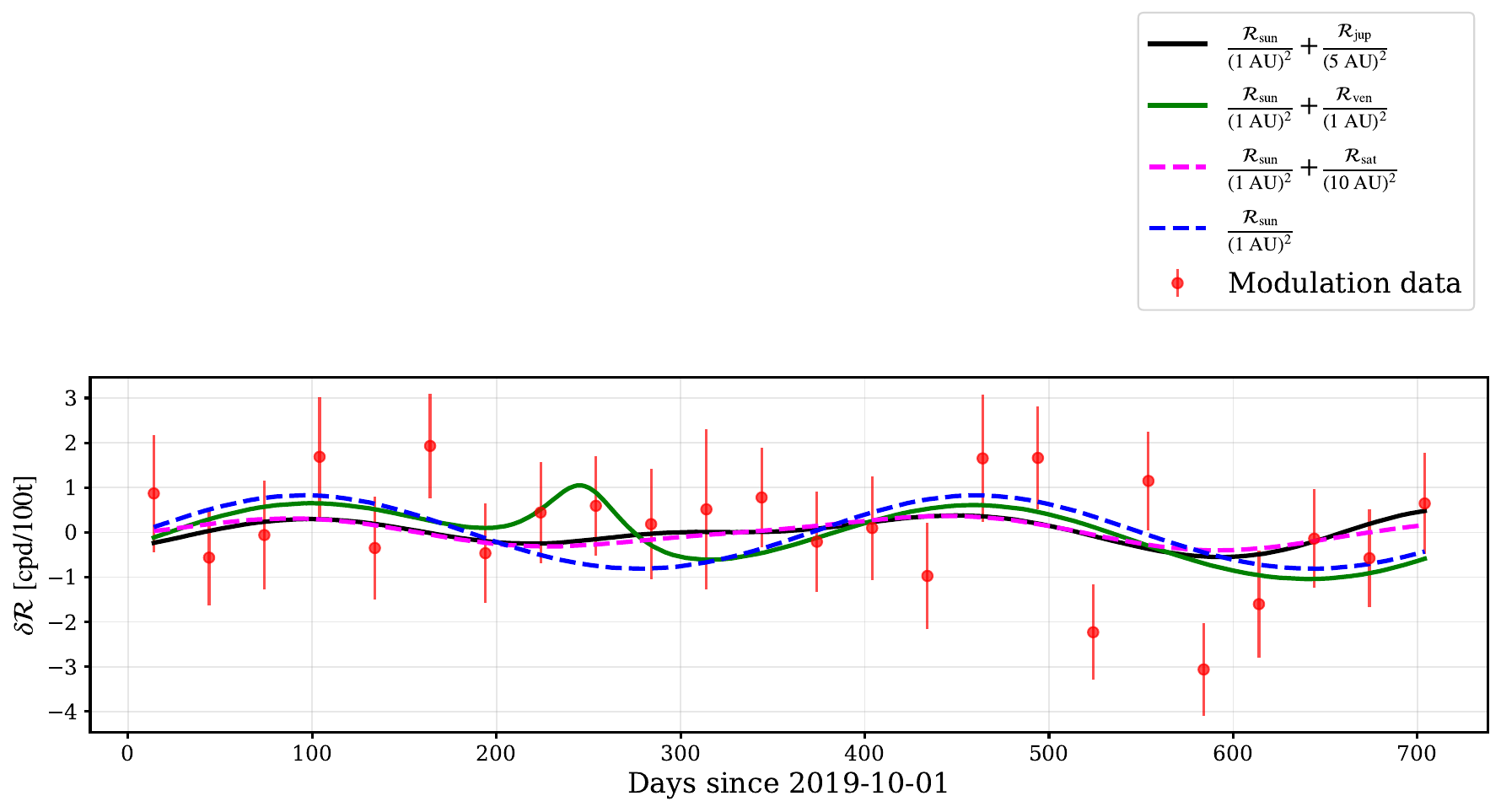}
    \caption{\label{fig:Modulationchi}
   Monthly-binned $\beta$-like event rate in the energy range corresponding to the $^7$Be solar neutrinos from October 2019 to October 2021 obtained by the BOREXINO collaboration. The black line shows the prediction with event rate contribution from Jupiter, the green line shows the prediction with event rate contribution from Venus, the magenta dashed line shows the prediction with event rate contribution from Saturn, and the blue dashed line shows the prediction with event rate contribution from only Sun using DE442s ephemeris
   $\frac{\chi^{2}_{sun+jup}}{dof}=\frac{23.6}{22},$
   $\frac{\chi^{2}_{sun+ven}}{dof}=\frac{24.0}{22},$
   $\frac{\chi^{2}_{sun+sat}}{dof}=\frac{24.7}{22}$
   $\frac{\chi^{2}_{sun}}{dof}=\frac{27.8}{23}$.}
\end{figure}

Similar to before, we now carry out $\chi^2$ goodness of fit tests from the monthly modulation data for the three hypotheses, for which the best-fit values are consistent with a positive signal. These include the models for which the monthly modulation  data consists of contributions from Sun, Sun+Venus, Sun+Jupiter and Sun+Saturn. We first superpose monthly modulation data on top of best-fits obtained for all the three models using Bayesian regression. This plot can be found in Fig.~\ref{fig:Modulationchi}. The reduced $\chi^2$ values which we get are 27.8/23 (1.2), 24.0/22, (1.1) 23.6/22 (1.07), 24.7/22 (1.1)  for Sun, Sun+Venus,  Sun+Jupiter, and Sun+Saturn,   respectively.
Therefore, the reduced $\chi^2$ values are close to 1, implying that all the  four models provide reasonable fits to the monthly modulation data.

\section{Results after using the full data}
\label{sec:fulldata}
\subsection {Analysis with rate time-series data}
\begin{figure}[h]
    \centering
    \includegraphics[width=0.6\textwidth]{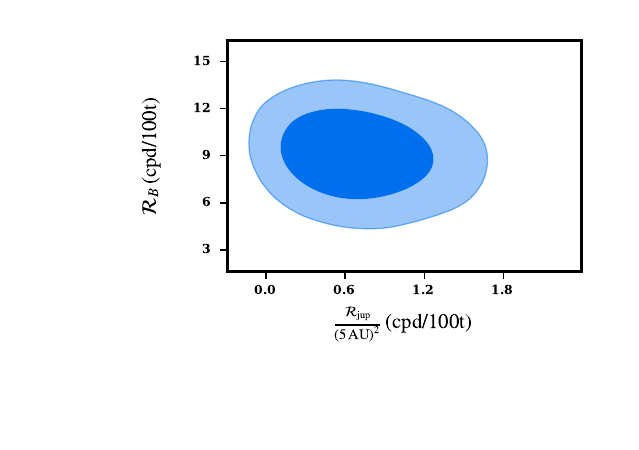}
    \caption{\label{fig:jup1full}
   \rthis{Marginalized 68\% and 95\% credible intervals for $\mathcal{R}_{\mathrm{B}}$ and  $\mathcal{R}_{\mathrm{jup}}$ after   normal prior on $\mathcal{R}_{\mathrm{sun}} \in \mathcal{N} (25,2)$ with units of (cpd/100t) using the full 10 years of BOREXINO data. The ephemerides used and the  priors on other parameters  are same as in Fig.~\ref{fig:jup1}. The marginalized 68\% value for $\frac{\mathcal{R}_{\mathrm{jup}}}{(5\,\mathrm{AU})^2}$ is given by \( 0.70^{+0.38}_{-0.40} \) (cpd/100t).} }
\end{figure}

\begin{figure}[h]
    \centering
    \includegraphics[width=0.9\textwidth]{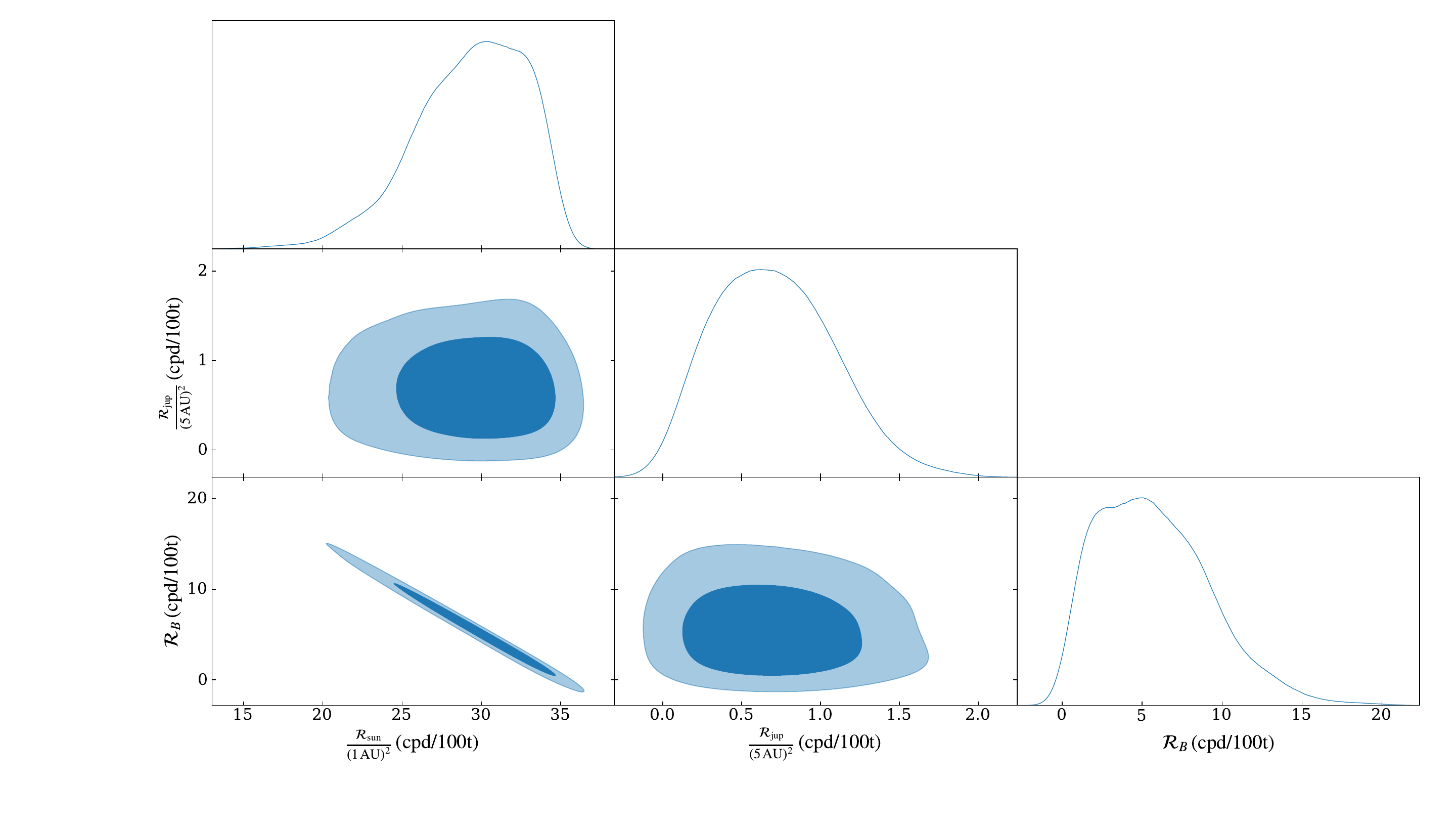}
    \caption{\label{fig:jup2full}
   \rthis{Marginalized 68\% and 95\% credible intervals for $\mathcal{R}_{\mathrm{sun}}$, $\mathcal{R}_{\mathrm{jup}}$ and $\mathcal{R}_{\mathrm{B}}$ using the full 10 years of BOREXINO data  after using  uniform prior on $\mathcal{R}_{\mathrm{sun}} \in \mathcal{U} [0, 50]$ with units of (cpd/100t). The ephemerides used and the  priors on other parameters  are the same as in Fig.~\ref{fig:jup1}.  The marginalized 68\% credible interval for $\frac{\mathcal{R}_{\mathrm{jup}}}{(5\,\mathrm{AU})^2}$ is given by $0.71^{+0.43}_{-0.38}$ (cpd/100t).}}
\end{figure}

\rthis{We now do the same analysis as in Sect.~\ref{sec:analysis} for Jupiter with the event rate time-series data over the full 10 year interval from December 2011 to October 2021. We carry out Bayesian regression on Eq.~\ref{eq:R1} using the same likelihood and priors as in Sect.~\ref{sec:analysis}.  The marginalized 68\% and 90\% credible intervals for  normal and uniform priors on $\mathcal{R}_{\mathrm{sun}}$ using the full data can be found in Fig.~\ref{fig:jup1full} and Fig.~\ref{fig:jup2full}, respectively for the DE422s ephemerides.
These best-fit values for \( \frac{\mathcal{R}_{\mathrm{jup}}}{(5\,\mathrm{AU})^2} \) are given by $0.70^{+0.38}_{-0.40}$ and $0.71^{+0.43}_{-0.38}$ (cpd/100t) respectively with significances of $1.75\sigma$ and $1.65\sigma$. The observed flux is about a factor of 2.2 smaller than that obtained using the analysis of 2019-2021 data. Furthermore, the significances are roughly comparable compared to using the data from 2019-2021. We also re-calculated the Bayes factors for the presence of an additional contribution of Jupiter compared to no contribution from Jupiter using the full data. We find Bayes factors of 1.1 and 1.4 for normal and uniform priors on  $\mathcal{R}_{\mathrm{sun}}$, respectively. Therefore, for both priors used, the Bayes factors are smaller than those used with the data between 2019-2021 and do not support any additional contribution from Jupiter.  We thereby conclude that the results from Bayesian model comparison do not support an additional contribution from Jupiter using the full rate time-series data.}

\subsection{Analysis with monthly modulation data}
\begin{figure}[htbp]
    \centering
    \includegraphics[width=0.5\textwidth]{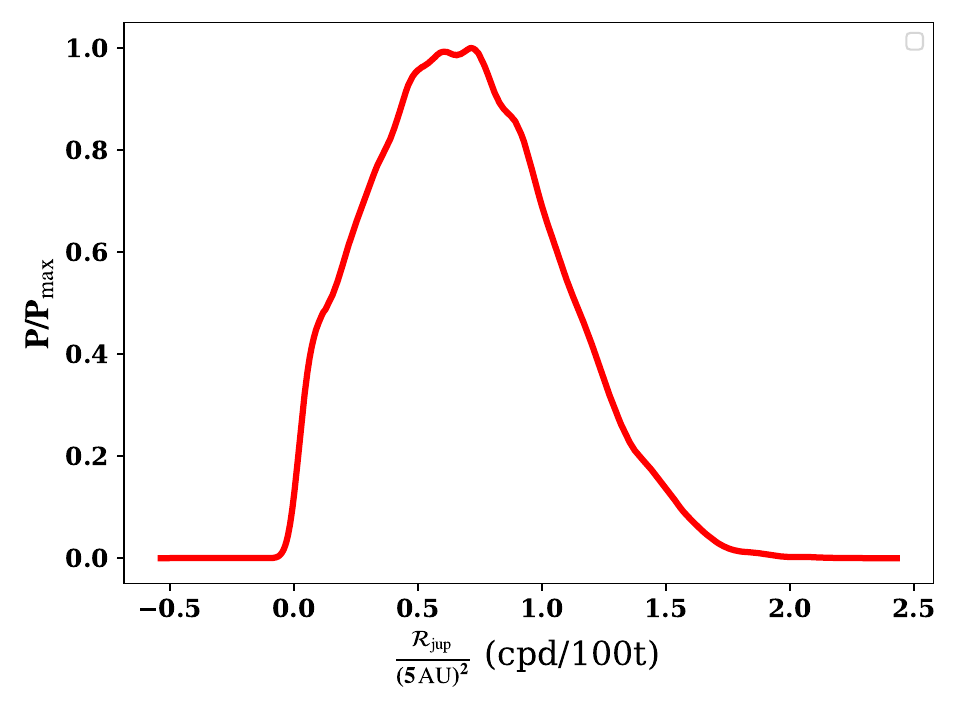}
    \caption{\label{fig:Modulationfull}
   \rthis{The marginalized posteriors on $\mathcal{R}_{\mathrm{jup}}$ for Jupiter  using the monthly modulation binned data (cf. Eq.~\ref{eq:4}) for the entire 10 years duration. The priors and ephemerides used are same as in Fig.~\ref{fig:Modulation}. The best-fit value for $\frac{\mathcal{R}_{\mathrm{jup}}}{(5\,\mathrm{AU})^2}$ for the full 10 years of data  is  equal to $0.65^{+0.42}_{-0.36}$ (cpd/100t). }}
   \end{figure}

\rthis{We now do the same analysis as in Sect.~\ref{sec:modulation}   using the full 10 years of data. We do a Bayesian regression analysis on Eq.~\ref{eq:4} using the same priors as in Sect.~~\ref{sec:modulation}. The best-fit marginalized values for $\mathcal{R}_{\mathrm{jup}}$ for the two time intervals can be found in Fig.~\ref{fig:Modulationfull} for the DE422s ephemerides.  The best-fit value for $\frac{\mathcal{R}_{\mathrm{jup}}}{(5\,\mathrm{AU})^2}$ for the full 10 years of data  is  equal to $0.65^{+0.42}_{-0.36}$ (cpd/100t) corresponding to  a significance  of $1.5\sigma$. 
The observed flux is about 3 times smaller than when using the data from 2019-21. Furthermore,  the significance is very slightly reduced compared to using the data from 2011-2013 or 2019-2021. We now calculate the Bayes factor (for an additional contribution from Jupiter) using the full modulation data with  the same priors as in Sect.~~\ref{sec:modulation}. We find the Bayes factor to be equal to 1.3. Therefore, the Bayes factor is reduced compared to that estimated in the intervals 2011-2013 and 2019-2021 and is consistent with no contribution from Jupiter.}

\section{Frequentist analyses of the data}
\label{sec:freq}
\subsection{Analysis with rate time-series}
\rthis{We now redo this search for a Jovian signal using the rate time-series data by carrying out a frequentist analysis to complement the Bayesian analysis done in the remainder of the manuscript. For this purpose,  we use profile likelihood to deal with the nuisance parameters.  More details about principle and application of profile likelihood to account to obtain frequentist confidence intervals can be found in ~\cite{PDG,Herold24} or some of our recent works~\cite{Vyaas,Barua25}. 
For analyzing the rate time-series data, we kept the solar contribution fixed to  $\mathcal{R}_{\mathrm{sun}}/ (1 \mathrm{AU})^2 = 25$ (cpd/100t). We then construct 
a grid of values for $\mathcal{R}_{\mathrm{jup}}$. For each value of
$\mathcal{R}_{\mathrm{jup}}$, we maximize the combined likelihood 
$\mathcal{L}(\mathcal{R}_{\mathrm{jup}},\mathcal{R}_{\mathrm{B}})$ with respect to  $\mathcal{R}_{\mathrm{B}}$
\begin{equation}
\mathcal{L}(\mathcal{R}_{\mathrm{jup}})= \max_{\mathcal{R}_{\mathrm{B}}}   
\mathcal{L}(\mathcal{R}_{\mathrm{jup}},\mathcal{R}_{\mathrm{B}})
\label{eq:pL}
\end{equation}
The central estimate for $\mathcal{R}_{\mathrm{jup}}$ can be obtained from
$\mathcal{L}(\mathcal{R}_{\mathrm{jup}})$. For this purpose,  we define 
$\chi^2 (\mathcal{R}_{\mathrm{jup}}) \equiv -2 \ln \mathcal{L}(\mathcal{R}_{\mathrm{jup}})$  and find its global minimum $\chi^2_{min}$.
We then obtain frequentist confidence intervals from $\Delta \chi^2$ distribution, where  $\Delta \chi^2 \equiv  \chi^2 (\mathcal{R}_{\mathrm{jup}}) - \chi^2_{min}$. For this purpose, we use Wilks' theorem, which states   that $\Delta \chi^2$ obeys a $\chi^2$ distribution for one degree of freedom~\cite{Wilks1938}.  The 68\% (1$\sigma$) confidence intervals can be obtained from the X-intercept for which $\Delta \chi^2=1$. All results for frequentist analysis have used the DE422s ephemerides. }

\begin{figure}[h]
    \centering
    \includegraphics[width=0.8\textwidth]{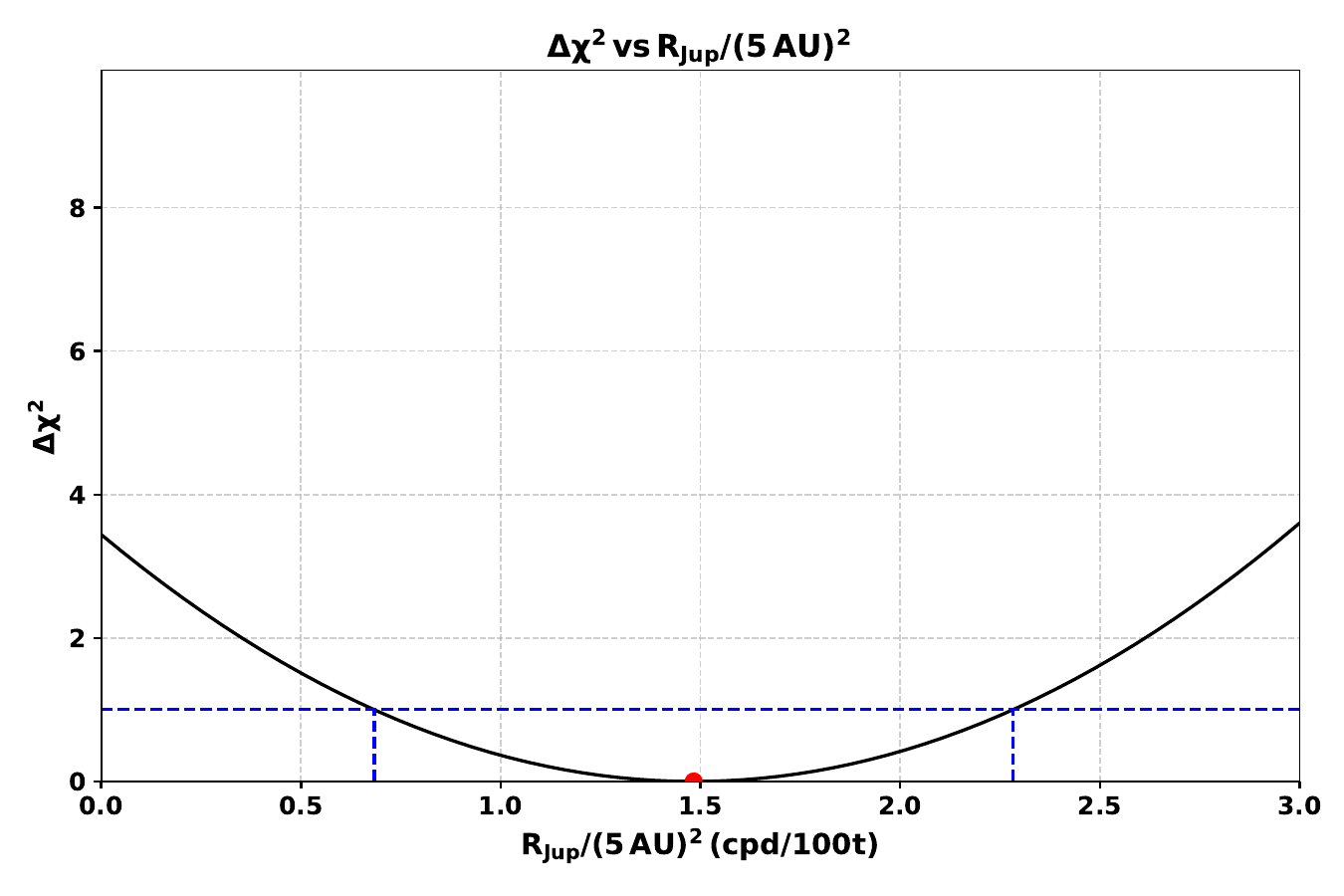}
    \caption{\label{fig:delta_chisq}
 \rthis{  $\Delta\chi^2$ as a function of $\frac{R_{\mathrm{jup}}}{(5\,\mathrm{AU})^2}$ using rate time series data for Oct. 2019 - Oct. 2021. The red dot corresponds to $\Delta \chi^2=0$ corresponding to the minimum value of $\chi^2$.  The dashed horizontal line corresponds to $\Delta\chi^2 = 1$ and the corresponding X-intercept is used  to obtain the 68\% ($1\sigma$)  confidence level estimates of $\frac{\mathcal{R}_{\mathrm{jup}}}{(5\,\mathrm{AU})^2}$. The best-fit value of $\mathcal{R}_{\mathrm{jup}}$ is given by $\frac{\mathcal{R}_{\mathrm{jup}}}{(5\,\mathrm{AU})^2}=1.48 \pm 0.8$ (cpd/100t).} }
\end{figure}

\rthis{The $\Delta \chi^2 $ for the rate time-series plot from October 2019-2021 can be found in Fig.~\ref{fig:delta_chisq}. We can see that $\Delta \chi^2$ shows a parabolic trend as a function of $\mathcal{R}_{\mathrm{jup}}$. The  central estimate for  $\mathcal{R}_{\mathrm{jup}}$ is given by $\frac{\mathcal{R}_{\mathrm{jup}}}{(5\,\mathrm{AU})^2}=1.48 \pm 0.80$ (cpd/100t). 
This  agrees with the Bayesian estimate within 1$\sigma$. The statistical significance of a non-zero flux from Jupiter according to this frequentist estimate is equal to $1.84\sigma$ and is marginally smaller than the Bayesian estimate of $2\sigma$.}

\begin{figure}[h]
    \centering
    \includegraphics[width=0.8\textwidth]{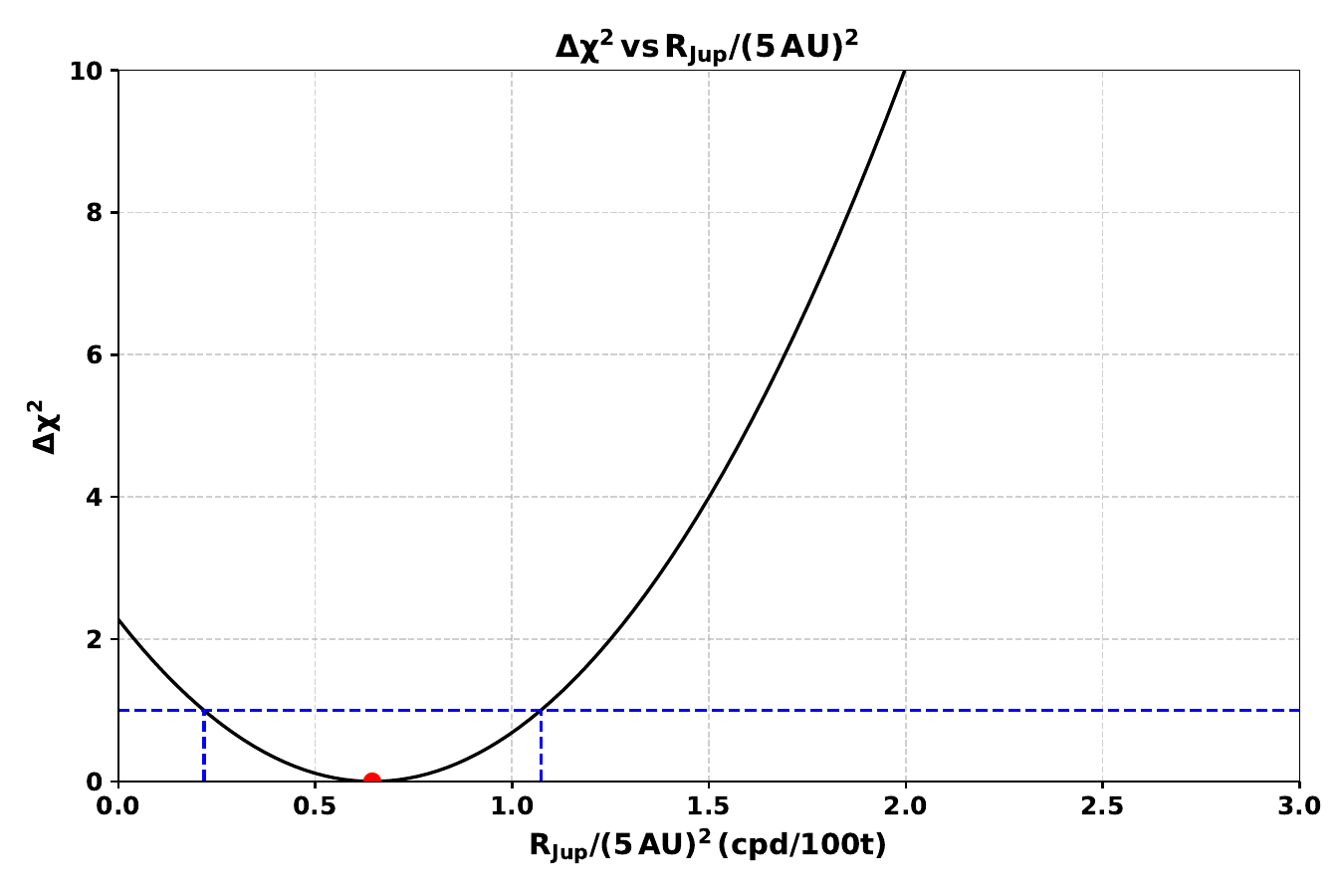}
    \caption{\label{fig:delta_chisq_alldata}
   \rthis{ $\Delta\chi^2$ as a function of $\frac{R_{\mathrm{jup}}}{(5\,\mathrm{AU})^2}$ using rate time series data for full 10 years of Borexino data. The plot is structured in the same way as Fig.~\ref{fig:delta_chisq}. The best-fit value of $\mathcal{R}_{\mathrm{jup}}$ is given by $\frac{\mathcal{R}_{\mathrm{jup}}}{(5\,\mathrm{AU})^2}=0.64 \pm 0.43$ (cpd/100t).}}
\end{figure}

\rthis{The corresponding plot using the full 10 years of  BOREXINO data can be found in Fig.~\ref{fig:delta_chisq_alldata}.  Once again, $\Delta \chi^2$ shows a parabolic trend as a function of $\mathcal{R}_{\mathrm{jup}}$ with a clear minimum.  The best-fit value for  $\mathcal{R}_{\mathrm{jup}}$ is given by $\frac{\mathcal{R}_{\mathrm{jup}}}{(5\,\mathrm{AU})^2}=0.64 \pm 0.43$ (cpd/100t), corresponding to a statistical significance of $1.6\sigma$.  This value and significance agree with the Bayesian estimate (cf.  Sect.~\ref{sec:fulldata}). Note however  that we do not get a central non-zero estimate at 90\% c.l., since the $\Delta \chi^2=2.71$ value  for the full dataset to the left of minimum occurs for $\mathcal{R}_{\mathrm{jup}}<0$. Therefore, we set a 90\% upper limit on 
$\frac{\mathcal{R}_{\mathrm{jup}}}{(5\,\mathrm{AU})^2}=<1.3$ (cpd/100t) using the full 10 years of data. }

\begin{figure}[htbp]
    \centering
    \includegraphics[width=0.9\textwidth]{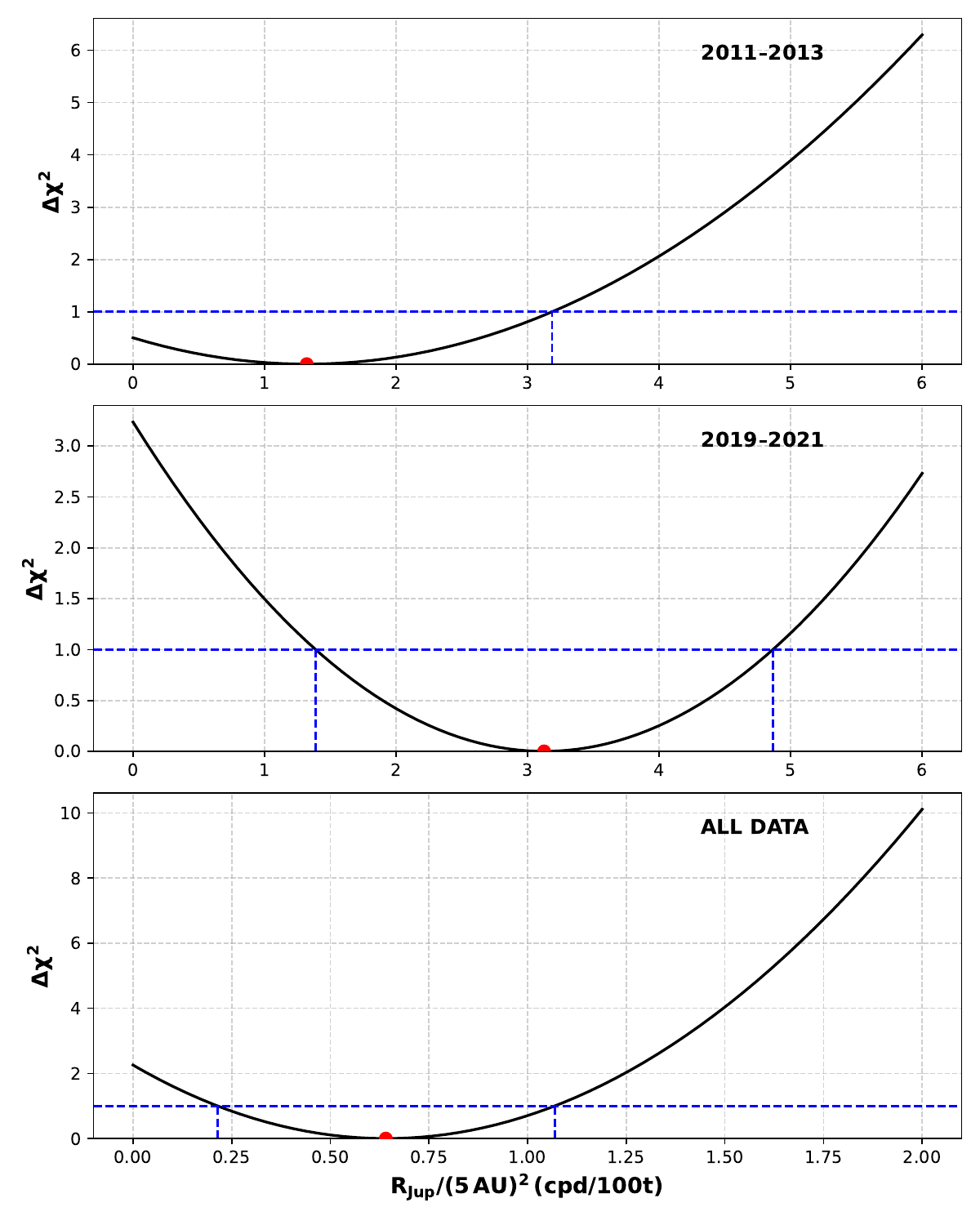}
    \caption{\label{fig:delta_chisq_modulation}
   \rthis{$\Delta\chi^2$ as a function of $\frac{R_{\mathrm{jup}}}{(5\,\mathrm{AU})^2}$ using the monthly modulation data for Dec. 2011 - Dec. 2013 (top panel), Oct. 2019 - Oct. 2021 (middle panel), and full 10 years of Borexino data (bottom panel). The plots in all the three panels are structured in the same way as Fig.~\ref{fig:delta_chisq}. The central estimate for $\frac{\mathcal{R}_{\mathrm{jup}}}{(5\,\mathrm{AU})^2}$ is equal to $3.12 \pm 1.74$ (cpd/100t) and  $0.64 \pm 0.43$ (cpd/100t), for 2019-2021 and for the full data, respectively, where the uncertainties refer to 68\% c.l. These central estimates have been obtained in the same way as in Fig.~\ref{fig:delta_chisq}. For 2011-13, we can only set an upper limit of  $\frac{\mathcal{R}_{\mathrm{jup}}}{(5\,\mathrm{AU})^2}<3.19 $ (cpd/100t) at 68\% c.l.}}
\end{figure}

\subsection{Analysis with monthly modulation data}
\rthis{We now  carry out a similar profile likelihood based frequentist analysis using the monthly modulation data. Since the background is already subtracted from the monthly modulation data, the only nuisance parameter involved is $\mathcal{R}_{\mathrm{sun}}$. Therefore, while analyzing the monthly modulation data, we consider of grid of values for $\mathcal{R}_{\mathrm{jup}}$ and maximize the likelihood (or minimize the $\chi^2$) with respect to $\mathcal{R}_{\mathrm{sun}}$ for each value of $\mathcal{R}_{\mathrm{jup}}$. We then find the minimum $\chi^2$ value over this grid,  and obtain a curve for   $\chi^2$ as a function of $\mathcal{R}_{\mathrm{jup}}$ similar to the rate time-series data. We do this analysis for three distinct periods: Oct. 2019-Oct. 2021, Dec. 2011-Dec. 2013,  and using the full 10 years of data. This plot for all the three periods can be found in Fig.~\ref{fig:delta_chisq_modulation}. For all the three time intervals, we see parabolic trend for $\Delta \chi^2$ as a function of  $\mathcal{R}_{\mathrm{jup}}$. However, for Dec. 2011-Dec. 2013 interval, the $\Delta \chi^2=1$ value corresponding   occurs at 
$\mathcal{R}_{\mathrm{jup}}<0$. Therefore, we can only set 68\% c.l. upper limits for $\mathcal{R}_{\mathrm{jup}}$ for the Dec. 2011-Dec. 2013 given by 
$\frac{\mathcal{R}_{\mathrm{jup}}}{(5\,\mathrm{AU})^2}<3.19 $ (cpd/100t). For 2019-2021 and the full 10 years BOREXINO data, we find that $\frac{\mathcal{R}_{\mathrm{jup}}}{(5\,\mathrm{AU})^2}$ is equal to  $3.12 \pm 1.74$ (cpd/100t) and  $0.64 \pm 0.43$ (cpd/100t), respectively where the uncertainties correspond to 68\% c.l. These correspond to significances of $1.8\sigma$ and $1.5\sigma$, respectively. Therefore, except for Dec. 2011- Dec. 2013, these values for  $\mathcal{R}_{\mathrm{jup}}$ agree with the Bayesian  credible intervals.}

\section{Conclusions}
\label{sec:conclusions}
In a recent work, A24 analyzed the BOREXINO $^7$Be solar neutrino data collected from 2011-2021. This data was found to deviate from the expected trend  of  $\propto \frac{1}{d_{\mathrm{sun}}^2}$. A24 posited that a flux contribution from Jupiter could
induce a signal similar to that of the $^7$Be signal with a time variation consistent with a $\frac{1}{d_{\mathrm{jup}}^2}$ dependence and also explain the discrepancy in the observed eccentricity.  A24 fit the total BOREXINO rate time series data between 2019-2021 using Bayesian regression to a contribution from Jupiter (in addition to Sun and the radioactive background) and found that Jupiter could contribute upto 6\% of the total signal with a significance of $\sim 2\sigma$. A24 then did a similar analysis using the monthly modulation data obtained after subtracting the known values of trend of the data from the rate time-series data between 2011-2013 and 2019-2021. They were able to confirm the previous result for a contribution from Jupiter during both these epochs. This signal from Jupiter was then argued to result from annihilation of dark matter particles of mass between 0.1-4 GeV trapped in the core of  Jupiter.

In this work, we have independently tried to reproduce this result in A24 \rthis{and also did multiple variants of their analysis}. We have also performed additional tests to ascertain the robustness of this signal. For this purpose, we did exactly a  similar search in both the rate time series data and modulation data for a signal from Venus, Mars, and Saturn, to see if a similar analysis shows a null result, since apriori no such signal should be expected from any other planet.  One slight difference in our analysis of the monthly modulation data is that we averaged over all the data (cf. Eq.~\ref{eq:4}) instead of an integral (cf. Eq.~\ref{eq:3}), as done in A24.
We also carried out hypothesis testing using Bayesian model comparison, by calculating the Bayes factor for an additional contribution from a given planet (Jupiter etc.) compared to the null hypothesis that the data only contain a contribution from the Sun (and possibly the radioactive background, if analyzing the total rate time-series data).  Finally, we also carried out $\chi^2$ goodness of fit tests using all the hypotheses considered. Our results are as follows:

\begin{itemize}
\item We were able to confirm using  Bayesian regression that the rate time series data for Oct. 2019 - Oct. 2021 as well as the monthly modulation data for Oct. 2019 - Oct. 2021 and Dec. 2011 - Dec. 2013 contain a nonzero contribution from Jupiter at about $2\sigma$ confidence level with flux of about 6\% that detected from the Sun.

\item We also confirmed that this result is consistent across different JPL ephemerides used to calculate the instantaneous distance to Jupiter.

\item When we  do a similar analysis for Mars, Venus, and Saturn,  we find that the signal from Mars is consistent with zero flux.

\item However, we see a non-zero signal from Venus in both the rate time series data for Oct. 2019 -  Oct. 2021 (for both the priors used) and in the modulation data for Oct. 2019 - Oct. 2021 with values roughly 10 times smaller than Jupiter at about $2\sigma$ significance level.

\item We also find a non-zero signal from Saturn between 2019-2021 in both the rate time series (when using a Gaussian prior on the solar flux contribution) and modulation data with significances of $1.6\sigma$ and  $1.8\sigma$, respectively and with flux about 1.7 times that of Jupiter.

\item When we carry our goodness of fit tests using both the rate time series and monthly modulation data, we find reduced $\chi^2$ values of close to one for all four hypotheses considered, namely that the data contain a contribution from only Sun, Sun+Jupiter, Sun+Venus, Sun+Saturn.

\item When we carry out Bayesian model comparison with both the rate time series and the modulation data, we find Bayes factor for an additional contribution from Jupiter  to be  less than 5. Similarly, the Bayes factors for additional contribution from Venus are less than one, indicating that it is not preferred compared to  a contribution from Sun. For Saturn, the Bayes factors are close to 1, implying that the evidence for an additional contribution from Saturn is marginal.

\item \rthis{When we do a Bayesian analysis using the full 10 years of data, the observed flux is about 2 (rate time-series)- 3 (monthly modulation) times smaller than that obtained using only the data from Oct. 2019 - Oct. 2021. However, the values of Bayes factors get reduced to around 1, implying that there is no evidence for additional contribution from Jupiter.}

\item \rthis{We also did a frequentist regression analysis for both the rate time series and monthly modulation data. The results mostly agree with those from Bayesian analysis, except for the monthly modulation data from Dec. 2011 - Dec. 2013, where we only get upper limits at 68\% c.l.} 
\end{itemize}

Therefore, we conclude that  even though Bayesian inference shows a flux from Jupiter, about 6\% of the $^7$Be flux during Oct. 2019 to Oct. 2021, Bayesian model comparison  provides only marginal evidence for this additional contribution from Jupiter. \rthis{This evidence is negligible when analyzing the full 10 years of BOREXINO data.} The Bayesian regression technique used in A24 also shows evidence for similar spurious contributions upto $2\sigma$  from Saturn and Venus, although once again  Bayesian model comparison does not provide any evidence for this contribution.  

Nevertheless, we agree with A24 that additional tests  should be done by the BOREXINO collaboration by comparing the direction of neutrino events with respect to Jupiter to obtain an additional degree of freedom. \rthis{Furthermore, it is also important to check that   the statistics of BOREXINO is sufficient to provide the declared sensitivity by anticipating the results with a dedicated Monte Carlo. These studies should be done by the BOREXINO Collaboration. As an additional test, one can also search for Jovian neutrinos using low energy (MeV energy range) data from the Super-Kamiokande experiment which is taking data since 1996.}

\begin{acknowledgments}
YHP has been supported by a Summer Undergraduate Research Exposure (SURE) Internship at IIT Hyderabad during summer of 2025. We are thankful to Saeed Ansarifard for patiently explaining all the nuances  of the analysis done in A24 and promptly answering all our queries. \rthis{We are also grateful to the anonymous referee for several constructive comments on our manuscript.}

\end{acknowledgments}

\bibliography{main}
\end{document}